\documentclass[11pt,a4paper]{article}
\usepackage{comment}
\usepackage{graphicx}
\usepackage{cancel}
\usepackage{amssymb}
\usepackage{amsmath}
\usepackage{amsfonts}
\usepackage{dsfont}
\usepackage{mathtools}
\usepackage{array}
\usepackage{soul}
\usepackage{rotating}
\usepackage{bbold,amsfonts}
\usepackage{breakurl}
\allowdisplaybreaks

\usepackage[utf8]{inputenc}
\usepackage{bm}
\usepackage{xcolor}
\usepackage{float}
\usepackage{braket}
\usepackage{placeins}
\usepackage[height=8.8in,width=6.45in]{geometry}
\usepackage[font=small,labelfont=bf]{caption}
\usepackage[hidelinks]{hyperref}
\usepackage{booktabs,float,slashed}
\usepackage{standalone}
\usepackage{caption}
\usepackage{subcaption}
\usepackage{makecell}
\usepackage[maxbibnames=99,sorting=none,giveninits=true,backend=biber,style=numeric-comp,sortcites,doi=false,hyperref=true]{biblatex}
\renewbibmacro{in:}{}
\usepackage{hyperref}
\DeclareFieldFormat{doilink}{\iffieldundef{doi}{#1}{\href{https://doi.org/\thefield{doi}}{#1}}}
\DeclareFieldFormat[article,periodical]{volume}{\mkbibbold{#1}}
\DeclareFieldFormat[article,periodical]{journaltitle}{#1}
\DeclareFieldFormat[article,periodical]{pages}{#1}

\usepackage{xpatch}
\xpatchbibdriver{article}
  {\usebibmacro{journal+issuetitle}%
   \newunit
   \usebibmacro{byeditor+others}%
   \newunit
   \usebibmacro{note+pages}}
  {\printtext[doilink]{%
     \usebibmacro{journal+issuetitle}%
     \newunit
     \usebibmacro{byeditor+others}%
     \newunit
     \usebibmacro{note+pages}}}
  {}{}

\numberwithin{equation}{section}

\newcommand{\beq}{\begin{equation}}
\newcommand{\eeq}{\end{equation}}

\newcommand{\overbar}[1]{\mkern 1.5mu\overline{\mkern-1.5mu#1\mkern-1.5mu}\mkern 1.5mu}

\makeatletter
\newcommand*{\letterdef@}{}
\newcommand*{\letterdef}[3]{%
	\def\letterdef@##1{\expandafter\newcommand\csname #1\endcsname{#2{##1}}}%
	\@tfor\@tempa :=#3\do{\expandafter\letterdef@\expandafter{\@tempa}}}
\makeatother
\letterdef{c#1} {\mathcal}{ABCDEFGHIJKLMNOPQRSTUVWXYZ} 
\letterdef{rm#1}{\mathrm} {dDeimM} 

\title{Integrated correlators with a Wilson line\\ in $\mathcal{N}=2$ SCFTs at strong coupling}

\begin{document}

\begin{titlepage}

\begin{flushright}
\small
\texttt{HU-EP-25/08}
\end{flushright}

\vspace*{10mm}
\begin{center}
{\LARGE \bf Integrated line-defect correlators\\[0.3em]  in $Sp(N)$ SCFTs at strong coupling
}

\vspace*{15mm}

{\Large L. De Lillo${}^{\,a,b}$,  M. Frau${}^{\,a,b}$ and  A. Pini${}^{\,c}$}

\vspace*{8mm}

${}^a$ Universit\`a di Torino, Dipartimento di Fisica,\\
			Via P. Giuria 1, I-10125 Torino, Italy
			\vskip 0.3cm
			
${}^b$   I.N.F.N. - sezione di Torino,\\
			Via P. Giuria 1, I-10125 Torino, Italy 
			\vskip 0.3cm

${}^c$ Institut f{\"u}r Physik, Humboldt-Universit{\"a}t zu Berlin,\\
     IRIS Geb{\"a}ude, Zum Großen Windkanal 2, 12489 Berlin, Germany  
     \vskip 0.3cm

\vskip 0.8cm
	{\small
		E-mail:
		\texttt{lorenzo.delillo,marialuisa.frau@unito.it;alessandro.pini@physik.hu-berlin.de}
	}
\vspace*{0.8cm}
\end{center}

\begin{abstract}
We consider two four-dimensional SCFTs with gauge group $Sp(N)$: the $\mathcal{N}=4$ SYM theory and the $\mathcal{N}=2$ theory with four hypermultiplets in the fundamental representation and one hypermultiplet in the rank-2 antisymmetric representation of the gauge group. Using supersymmetric localization and by exploiting a Toda equation, we compute the integrated correlator between a half-BPS Wilson line and two moment map operators of conformal dimension 2. We obtain exact expressions for this observable, valid for any value of the 't Hooft coupling in the large $N$-limit of the theory and we derive the corresponding strong coupling expansions.
\end{abstract}
\end{titlepage}
\setcounter{tocdepth}{2}

\newpage

\tableofcontents

\vspace*{1cm}

\section{Introduction}
The study of integrated correlators in four dimensional gauge theories has attracted increasing attention during recent years, as they constitute a useful tool to explore the strong coupling regime
and the non-perturbative aspects of the theory.
In particular, these observables have been deeply studied in the context of $\mathcal{N}=4$ SYM, the maximally supersymmetric theory in four dimensions, where the integrated correlator among four local scalar operators belonging to the stress-tensor multiplet has been considered.
It is important to recall that while the space-time dependence of 2- and 3-point functions is fully determined by conformal symmetry, the case of 4-point functions requires more attention, as their space-time dependence can only be reduced to a single function of two conformal invariant cross-ratios \cite{Dolan:2000ut}. Nevertheless, we can still extract a lot of information from this last correlator by integrating out the dependence on the space-time coordinates with an appropriate integration measure that is completely determined by superconformal invariance.
This approach was initiated in \cite{Binder:2019jwn} where it was shown that the evaluation of this observable can be efficiently performed by exploiting supersymmetric localization \cite{Pestun:2007rz}. This amounts to place the $\mathcal{N}=4$ SYM theory on a unit 4-sphere $\mathbb{S}^4$, and deform it to the so-called $\mathcal{N}=2^{\star}$ theory, where the adjoint hypermultiplet acquires a mass $m$. Then, by taking four derivatives of the logarithm of the mass deformed partition function $\mathcal{Z}_{\mathcal{N}=2^{\star}}$ one obtains
\begin{subequations}
\begin{align}
& \partial_{\tau_p}\partial_{\bar{\tau}_p}\partial_{m}^2\log \mathcal{Z}_{\mathcal{N}=2^{\star}}\Big|_{m=0} = \int \prod_{i=1}^{4}dx_i\, \mu(x_i)\, \langle \mathcal{O}_p(x_1)\overbar{\mathcal{O}}_p(x_2)\mathcal{O}_2(x_3)\overbar{\mathcal{O}}_2(x_4) \rangle_{\mathcal{N}=4} \, \ , \\
& \partial_{m}^4\log \mathcal{Z}_{\mathcal{N}=2^{\star}}\Big|_{m=0} = \int \prod_{i=1}^{4}dx_i\, \mu'(x_i)\, \langle \mathcal{O}_2(x_1)\overbar{\mathcal{O}}_2(x_2)\mathcal{O}_2(x_3)\overbar{\mathcal{O}}_2(x_4) \rangle_{\mathcal{N}=4}\, \ ,
\end{align}
\label{IntegratedFirstType}%
\end{subequations}
where $\mathcal{O}_{p}(x_i), \overbar{\mathcal{O}}_{p}(x_i)$ denote 1/2 BPS chiral primary operators with conformal dimension $p$, transforming in the $[0,p,0]$ representation of the $SU(4)_R$ R-symmetry group. Furthermore, $\mu(x_i)$ and $\mu'(x_i)$ are the two integration measures that are completely determined by superconformal invariance and finally $\tau_p$ and $\overbar{\tau}_p$ represent the couplings associated with the  $\mathcal{O}_p(x)$ and $\overbar{\mathcal{O}}_{p}(x)$ operators, respectively \cite{Gerchkovitz:2016gxx}. Many properties of the correlators \eqref{IntegratedFirstType} have been extensively studied in recent years. These include their weak coupling expansions \cite{Wen:2022oky,Zhang:2024ypu}, achieved through explicit Feynman diagram computations that validate the results from localization, as well as their modular properties \cite{Chester:2019jas,Dorigoni:2021bvj,Dorigoni:2021guq,Chester:2020vyz,Chester:2019pvm,Alday:2021vfb,Dorigoni:2022cua,Collier:2022emf,Paul:2022piq,Alday:2023pet,Dorigoni:2024dhy,Kim:2024pjb,Chester:2020dja}. Extensions to generic gauge groups have also been analyzed \cite{Dorigoni:2022zcr,Dorigoni:2023ezg}. Cases involving the insertion of operators with large charge \cite{Paul:2023rka,Brown:2023why,Brown:2023cpz}, operators with arbitrary conformal dimensions \cite{Brown:2023zbr} as well as determinant operators \cite{Brown:2024tru} have also been explored. Lastly, some of these findings have been partially extended to $\mathcal{N}=2$ SCFTs \cite{Chester:2022sqb,Fiol:2023cml,Behan:2023fqq,Billo:2023kak,Pini:2024uia,Billo:2024ftq}.

Until very recently, most of the attention has been directed toward the study of the integrated correlators \eqref{IntegratedFirstType}, which involve only the insertion of local operators. However, integrated correlators with the insertion of defect operators can also be considered. In particular, in the context of $\mathcal{N}=4$ SYM, one can examine the correlator between a half-BPS Wilson line and two chiral operators $\mathcal{O}_{2}(x_i)$. By exploiting conformal invariance, the space-time dependence of this correlator can be expressed as a function of two conformally invariant cross-ratios and the coupling constant \cite{Buchbinder:2012vr}. Moreover, in this case as well, the dependence on the space-time coordinates can be integrated out using an integration measure fixed by superconformal invariance \cite{Billo:2023ncz,Billo:2024kri,Dempsey:2024vkf}, thus obtaining the corresponding integrated correlator. Also this observable can be computed using supersymmetric localization. Specifically, it can be obtained by taking two derivatives of the logarithm of the Wilson loop vacuum expectation value $\langle W \rangle$ in the mass-deformed $\mathcal{N}=2^{\star}$ theory, namely \cite{Pufu:2023vwo}
\begin{align}
\label{Wint}
\partial_{m}^2 \log \langle W \rangle \big|_{m=0} = \int d^4x_1 \, d^4x_2 \, \hat{\mu}(x_1, x_2) \, \langle \mathcal{O}_2(x_1) \overbar{\mathcal{O}}_2(x_2) \rangle_{W} \, ,
\end{align}
where $\langle \mathcal{O}_2(x_1) \overbar{\mathcal{O}}_2(x_2) \rangle_W$ denotes the 2-point function of the chiral operators in the presence of a Wilson line, while $\hat{\mu}(x_1,x_2)$  refers to the integration measure mentioned above. 
This second type of integrated correlator is not only relevant for constraining bootstrap computations (see e.g. \cite{Chester:2022sqb}), but its importance also
lies in the fact that, since line operators generally transform non-trivially under the $\mathcal{S}$-duality group \cite{MONTONEN1977117} of $\mathcal{N}=4$ SYM, they provide an observable well suited for studying the implications of this symmetry. Using this feature, the modular properties of the correlator \eqref{Wint}, as well as its generalization involving the insertion of dyonic operators, have been analyzed in \cite{Pufu:2023vwo,Dorigoni:2024vrb,Dorigoni:2024csx}.

Finally, as argued in \cite{Dempsey:2024vkf}, integrated correlators involving the insertion of line defect operators can also be considered in the context of $\mathcal{N}=2$ SCFTs. These can still be computed using supersymmetric localization and the analogous of the relation \eqref{Wint}, where the chiral operators $\mathcal{O}_2(x_i)$ of the $\mathcal{N}=4$ SYM theory are replaced by two moment map operators of conformal dimension 2. These 
are superconformal primary operators, arising as the top component of the short multiplet $\mathcal{\hat{B}}_1$ of the $\mathfrak{su}(2,2|2)$ superconformal algebra \cite{Dolan:2002zh}. However, so far, the only explicit computations for non-maximally supersymmetric theories concern the planar and next-to-planar terms of the integrated correlator \eqref{Wint} for the two-node $\mathcal{N}=2$ quiver gauge theory, which arises as a $\mathbb{Z}_2$ orbifold of $\mathcal{N}=4$ SYM \cite{Pini:2024zwi}.

In this work, we aim to extend the study of the integrated correlators \eqref{Wint} to theories with gauge group $Sp(N)$ \footnote{In this paper we employ the same notation of \cite{Billo:2024ftq}, where $Sp(N)$ indicates the symplectic group of rank $N$ .}. Specifically, we focus on the $\mathcal{N}=4$ SYM theory and the $\mathcal{N}=2$ theory, whose matter content consists of four hypermultiplets in the fundamental representation and one hypermultiplet in the rank-2 antisymmetric representation of $Sp(N)$. This theory is planar-equivalent to $\mathcal{N}=4$ SYM, as demonstrated, for example, by the large-$N$ evaluations of the free energy and the v.e.v. of the $\frac{1}{2}$-BPS circular Wilson loop carried out in \cite{Beccaria:2022kxy,Beccaria:2021ism}. Furthermore it admits a gravity dual of the form $AdS_5\times \mathbb{S}^5/\mathbb{Z}_2$, obtained through a proper $\mathbb{Z}_2$ orientifold procedure applied to the holographic dual of $\mathcal{N}=4$ SYM \cite{Ennes:2000fu}. Consequently, our results can be used to formulate predictions for holographic computations. An additional motivation for studying these theories lies in the fact that, although the correlator \eqref{Wint} has been thoroughly analyzed for theories based on the $SU(N)$ gauge group, the case of other types of Lie groups has received no attention so far. This work represents the first step in that direction. Moreover, we also extend the set of $\mathcal{N}=2$ SCFTs for which exact expressions for this correlator can be obtained in the large-$N$ limit, as the only other case discussed in the literature is that of \cite{Pini:2024zwi}. The lack of results for this observable in non-maximally supersymmetric theories stems from the fact that applying supersymmetric localization in the context of $\mathcal{N}=2$ theories leads to a matrix model with a non-trivial potential, unlike the case of $\mathcal{N}=4$ SYM, where the matrix model is  Gaussian. This, in turn, makes the derivation of exact expressions particularly challenging, especially in the strong coupling regime. To overcome these difficulties, we develop a novel computational method to obtain the large-$N$ expansion of this correlator, which differs from the one of \cite{Pufu:2023vwo} that is based on topological recursions \cite{Eynard:2004mh}. Our method relies on a Toda-like equation, which holds in both theories considered in this work, and whose existence and main properties were previously discussed in \cite{Beccaria:2022kxy}. This Toda equation establishes a relation between quantities in the $Sp(N)$ theory and those in the $Sp(N\pm1)$ theories, which allows us to recursively determine the coefficients of the large-$N$ expansions.
Remarkably, using this method, we are able to evaluate the correlator \eqref{Wint} up to a very high order.

The rest of this work is organized as follows. In Section \ref{sec:MatrixModel}, we review the main properties of the two gauge theories considered in this paper. In particular, we discuss the key features of the corresponding matrix models and explain how the computation of the integrated correlator \eqref{Wint} can be performed using supersymmetric localization. Then, in Section \ref{sec:TodaChain}, we review the Toda equation, which constitutes our main computational tool. In particular, we explain how it can potentially be  used to obtain the large-$N$ expansions of a large class of observables. In Section \ref{sec:I4dN2} we evaluate the large-$N$ expansion of the integrated correlator \eqref{Wint} for the above mentioned $\mathcal{N}=2$ theory. In Section \ref{sec:IN4}, using the $\mathcal{N}=2^{\star}$ mass-deformed theory, we perform the analogous computation for the $\mathcal{N}=4$ SYM theory with $Sp(N)$ gauge group. In both cases, we derive exact expressions valid for any value of the 't Hooft coupling, which can subsequently be expanded at strong coupling. Our conclusions are presented in Section \ref{sec:Conclusions}. Finally, the proofs of the mathematical identities used in the derivation of our results and technical details concerning the mathematical techniques employed in deriving the strong coupling expansions have been collected in Appendix \ref{app:identities} and \ref{SC} respectively. In Appendix \ref{App:Details on N=4 SYM-like theory}, we furnish further details regarding how to obtain the large-$N$ expansions in the Gaussian theory.

\section{\texorpdfstring{$Sp(N)$}{} theories and their mass deformations}
\label{sec:MatrixModel}
The first theory we consider is $\mathcal{N}=4$ SYM with gauge group $Sp(N)$.
The second one is the $\mathcal{N}=2$ theory realized on a stack of $2N$ D3-branes in presence of eight D7-branes and a single O7 orientifold plane \cite{Aharony:1996en,Ennes:2000fu}. This gauge theory has four hypermultiplets in the fundamental representation and one hypermultiplet in the rank-2 antisymmetric representation of $Sp(N)$. In the following we will often refer to this theory as the ``$\mathcal{N}=2$ model''.
A summary of the matter content of these two theories is provided in Table \ref{tab:MatterContent}. It can be verified that in both cases the $\beta$-functions vanish, making these theories conformal.
\begin{table}[h!]
\centering
\begin{tabular}{c|c|c|c|} 
& $n_{\text{adj}}$ & $n_{\text{F}}$ & $n_{\text{A}}$ \\ \hline\hline
$\mathcal{N}=4$ SYM & 1 & 0 & 0 \\ \hline
$\mathcal{N}=2$ Theory & 0 & 4 & 1 \\ \hline
\end{tabular}
\caption{The two superconformal theories with group $Sp(N)$.}
\label{tab:MatterContent}
\end{table}

By placing the gauge theory on a unit 4-sphere $\mathbb{S}^4$ and using supersymmetric localization \cite{Pestun:2007rz}, we can reduce the computation of the partition function $\mathcal{Z}$ to the evaluation of the following expression
\begin{align}
\mathcal{Z}=  {\mathcal C}_N 
\int da\, \, \text{e}^{-\frac{8\pi^2N}{\lambda}\text{tr}\,a^2}\, \Big|{Z}_{\text{1-loop}}(a)\,{Z}_{\text{inst}}(a)\Big|^2\, \ ,
\label{Z}
\end{align}
where ${\mathcal C}_N$ is a normalization constant, $\lambda = g^2\,N$ is the 't Hooft coupling, and $a$ is a matrix belonging to the $\mathfrak{sp}(N)$ Lie algebra\footnote{
We employ the following conventions
\begin{align}
a = \sum_{b=1}^{N(2N+1)}a_bT^{b}\, ,\qquad \qquad  da = \prod_{b=1}^{N(2N+1)}\frac{d\,a_b}{\sqrt{2\pi}}\, \ ,    
\end{align}
where $T^{b}$ denote the generators of the $\mathfrak{sp}(N)$ Lie algebra in the fundamental representation, normalized such that $\text{tr}\,T^{a}T^{b}=\frac{1}{2}\delta^{a,b}$
.}. Finally ${Z}_{\text{1-loop}}(a)$ and ${Z}_{\text{inst}}$(a) represent the 1-loop and instanton contributions respectively. In the large-$N$ limit where instantons are exponentially suppressed, we can set  ${Z}_{\text{inst}}(a)=1$. On the other hand, the 1-loop contribution depends on the  matter content of the theory and reads \cite{Fiol:2020bhf}
\begin{align}
\Big|{Z}_{\text{1-loop}}(a)\Big|^2 = \frac{ \displaystyle \prod_{i<j}^{N}\Big[H(a^{+}_{ij})\Big]^2\Big[H(a_{ij}^{-})\Big]^2\prod_{i=1}^{N}\Big[H(2a_i)\Big]^2}{ \displaystyle \prod_{i<j}^{N}\Big[H(a_{ij}^{+})\Big]^{2n_{\text{adj}}+2n_{\text{A}}} \Big[H(a_{ij}^{-})\Big]^{2n_{\text{adj}}+2n_\text{A}}\prod_{i=1}^{N}\Big[H(2a_i)\Big]^{2n_{\text{adj}}}\Big[H(a_i)\Big]^{2n_\text{F}}}\, \ ,
\label{Z1loop}
\end{align}
where $a_{ij}^{\pm} = a_i \pm a_j$ and the function $H(x)$ is defined as
\begin{align}
H(x) = \text{e}^{-(1+\gamma)x^2}G(1+ix)G(1-ix)\,
\label{H(x)}
\end{align}
with $\gamma$ being the Euler-Mascheroni constant and $G$ the Barnes G-function.

To compute the integrated correlator \eqref{Wint} we need to introduce a mass deformation. In the case of $\mathcal{N}=4$ SYM we give a mass $m$ to the adjoint hypermultiplet obtaining the so called $\mathcal{N}=2^{\star}$ theory while, for the $\mathcal{N}=2$ model, we assign masses $m_f$ (with $f=1,\dots,4$ ) to the fundamental hypermultiplets. In both cases, this deformation gives rise to a mass-dependent partition function which, for the purposes of this work, we need to consider only up  to order $O(m^2)$. For example, for the $\mathcal{N}=2^{\star}$ theory we have
\begin{align}
\mathcal{Z}_{\mathcal{N}=2^{\star}}(m) ={\mathcal C}_N 
\int da\, \text{e}^{-\frac{8\pi^2N}{\lambda}\text{tr}\,a^2}\, \Big|{Z}^{\mathcal{N}=2^{\star}}_{\text{1-loop}}(a,m)\Big|^2\, \ ,
\label{ZN2star}
\end{align}
where
\begin{align}
& \Big|{Z}^{\mathcal{N}=2^{\star}}_{\text{1-loop}}(a,m)\Big|^2 = \prod_{i=1}^{N}\frac{\Big[H(2a_i)\Big]^2}{H(2a_i+m)H(2a_i-m)}\prod_{i<j}^{N}\frac{\Big[H(a^{+}_{ij})\Big]^2\Big[H(a_{ij}^{-})\Big]^2}{H(a^{+}_{ij}+m)H(a^{+}_{ij}-m)H(a^{-}_{ij}+m)H(a^{-}_{ij}-m)} \nonumber \\
& = \exp\left[1-m^2\sum_{i=1}^{N}\partial^2\log H(2a)-m^2\sum_{i<j}^{N}\left(\partial^2\log H(a_{ij}^{+})+\partial^2\log H(a_{ij}^{-})\right) + O(m^4)\right]\, \ .  
\end{align}
If we rescale the matrix $a$ as follows
\begin{align}
a \mapsto \sqrt{\frac{\lambda}{8\pi^2N}}\,a    
\label{ar}
\end{align}
we can employ the expansion
\begin{align}
\log H(x) = \sum_{n=1}^{\infty}(-1)^n\frac{\zeta_{2n+1}}{n+1}x^{2n+2}\,  ,   
\end{align}
where $\zeta_k$ denotes the value of the Riemann zeta function $\zeta(k)$, to obtain the small-mass expansion of the partition function \eqref{ZN2star}. This reads
\begin{align}
\mathcal{Z}_{\mathcal{N}=2^{\star}}(m) = {\mathcal C}_N \, \Big(  \frac{\lambda}{8\pi^2N} \Big)^{\frac{N(2N+1)}{2}} 
\int da \,\text{exp}\left[\displaystyle -\text{tr}a^2+m^2\,\widetilde{M}+O(m^4)\right]\, \ ,
\label{ZN2starExpansion}
\end{align}
where
\begin{align}
\widetilde{M} =  \widetilde{M}^{(1)} + {\widetilde{M}}^{(2)}\, \ 
\label{MN2star}
\end{align}
\label{SmallMassExpansionZSpData}
with
\begin{subequations}
\begin{align}
& \widetilde{M}^{(1)} =  -\frac{1}{2}\sum_{n=1}^{\infty}(-1)^n(2n+1)\zeta_{2n+1}\Big(\frac{\lambda}{2\pi^2N}\Big)^{n}\text{tr}\,a^{2n} \, , \\[0.5em]
& \widetilde{M}^{(2)} = -\frac{1}{2}\sum_{n=1}^{\infty}\sum_{k=0}^{n}(-1)^n\zeta_{2n+1}(2n+1)\Big(\frac{\lambda}{8\pi^2N}\Big)^{n}\binom{2n}{2k}
\,\text{tr}\,a^{2n-2k}\,\text{tr}\,a^{2k}\, \ .
\end{align}
\label{M1andM2}%
\end{subequations}
By following the same steps we can derive the small-mass expansion for the partition function of the $\mathcal{N}=2$ model which is given by
\begin{align}
\mathcal{Z}(m_f) = {\mathcal C}_N \, \Big(  \frac{\lambda}{8\pi^2N} \Big)^{\frac{N(2N+1)}{2}} 
\int da \,\text{exp}\bigg[-\text{tr}\,a^2-S_0+\sum_{f=1}^{4}m_f^2\,{M}+O(m_f^4)\,\bigg]\, \ ,
\label{Zmf}
\end{align}
where
\begin{align}
& S_0 = 4\sum_{n=1}^{\infty}(-1)^{n+1}\Big(\frac{\lambda}{8\pi^2N}\Big)^{n+1}(2^{2n}-1)\,\frac{\zeta_{2n+1}}{n+1}\,\text{tr}\,a^{2n+2}\,\  ,
\label{SOtilde}\\[0.5em]
& {M} = -\sum_{n=1}^{\infty}(-1)^n(2n+1)\,\zeta_{2n+1}\Big(\frac{\lambda}{8\pi^2N}\Big)^n\,\text{tr}\,a^{2n}\, \ .
\label{Msp}    
\end{align}
It is worth noting that, apart from the presence of the interaction action $S_0$, the difference between \eqref{ZN2starExpansion} and \eqref{Zmf} lies in the distinct forms of the $m^2$-terms, namely the operators \eqref{MN2star} and \eqref{Msp}. In particular, the $\widetilde{M}$ operator is more involved, as it also includes a double-trace contribution.

Next, we recall that the half-BPS circular Wilson loop in the fundamental representation admits \cite{Pestun:2007rz, Beccaria:2021ism} a simple matrix-model description in terms of the rescaled matrix $a$ \eqref{ar} given by \footnote{The Wilson loop is usually normalized by dividing the expression  \eqref{WilsonLoopSp} by the dimension of the fundamental representation of $Sp(N)$, which is $2N$. However, this overall factor does not influence the evaluation of the integrated correlator \eqref{Wint}. For simplicity, we will therefore omit this extra factor.
} 
\begin{align}
W(a,\lambda) = 
\sum_{k=0}^{\infty}\frac{1}{k!}\Big(\frac{\lambda}{2N}\Big)^{\frac{k}{2}} \text{tr}a^{k} 
\label{WilsonLoopSp}   
\end{align}
so that its v.e.v. can be easily computed in both mass-deformed theories. In the $\mathcal{N}=2^{\star}$ theory we have
\begin{align}
\label{WvevN2star}
& \widetilde{\mathcal{W}}(m,\lambda) \equiv  \Big( \frac{\lambda}{8\pi^2N} \Big)^{\frac{N(2N+1)}{2}} \, 
\frac{{\mathcal C}_N}{\mathcal{Z}_{\mathcal{N}=2^{\star}}(m)}\int da \, W(a,\lambda) \,\text{exp}\left[\displaystyle -\text{tr}a^2+m^2\, \widetilde{M}+O(m^4)\right]\, \ .  
\end{align}    
Thus the integrated correlator \eqref{Wint} for $\mathcal{N}=4$ SYM is given by
\begin{align}
& \widetilde{\mathcal{I}}(\lambda) \equiv \partial_{m}^2\log \widetilde{\mathcal{W}}(m,\lambda)\Big|_{m=0} = 2\,\frac{\langle W\,\widetilde{M} \rangle_{0} - \langle W \rangle_{0}\,\langle\,  \widetilde{M}\rangle_{0}}{\langle W \rangle_{0}} \, ,
\label{ISpN4}
\end{align}
where the subscript $_0$ in the right-hand side means that the v.e.v.'s are computed in the
Gaussian matrix model.
In a similar way one finds that the integrated correlator \eqref{Wint} for the $\mathcal{N}=2$ model is
\begin{align}
\mathcal{I}(\lambda) \equiv \partial_{m_f}^2\log \mathcal{W}(m_f,\lambda)\Big|_{m_f=0} = 2\frac{\langle W\,{M} \rangle - \langle W \rangle\,\langle {M}\rangle}{\langle W\rangle}\, \ ,
\label{ISpN2}    
\end{align}
where the v.e.v.'s in the right-hand side are computed in the  matrix model corresponding to the $\mathcal{N}=2$ theory.  For a generic function $f(a)$ such a v.e.v. is given by
\begin{align}
\langle f(a) \rangle = \frac{\langle f(a)\,\text{e}^{-S_0} \rangle_0}{\langle \text{e}^{-S_0} \rangle_0} \, \,
\end{align}
and thus also in this case everything is reduced to a calculation in the Gaussian matrix model.

The v.e.v. in \eqref{ISpN4} and \eqref{ISpN2} can in general be computed by starting from  the fusion/fission relations \cite{PhysRevD.14.1536,Huang:2016iqf} and the set of identities satisfied by the vacuum expectation values of multiple traces of the rescaled $a$ matrix  \cite{Beccaria:2020hgy,Beccaria:2021ism}
\begin{align}
t_{p_1,\cdots,p_M} \equiv \langle \text{tr}\,a^{\,p_1}\,\cdots\,\text{tr}\,a^{\,p_M} \rangle_0\, \ .
\label{tfunctions}
\end{align}
As shown in \cite{Billo:2024ftq}\footnote{The same method has  been successfully applied also in the context of gauge theories with $SU(N)$ gauge group, see e.g. \cite{Beccaria:2020hgy,Billo:2022fnb,Billo:2022lrv,Beccaria:2021hvt}
.} the calculation of expectation values in the interacting theory can be considerably simplified by performing the following linear change of basis
\begin{align}
\mathcal{P}_k = \sqrt{\frac{k}{2}}\,\sum_{\ell=0}^{[\frac{k-1}{2}]} (-1)^\ell \Big( \frac{N}{2} \Big)^{\ell-\frac{k}{2}} \frac{(k-\ell-1)!}{\ell!(k-2\ell)!} \left( \text{tr}\,a^{k-2\ell} - \langle \text{tr}\,a^{k-2\ell} \rangle_0  \right) \;,
\label{PK}    
\end{align}
where the operators $\mathcal{P}_k$ are orthonormal in the Gaussian matrix model in the planar limit:
\begin{equation}
    \langle \mathcal{P}_{2k_1}\mathcal{P}_{2k_2} \rangle_0 = \delta_{k_1,k_2} + O \Big( \frac{1}{N}\Big) \;.
    \label{Orthonormality}
\end{equation}

As discussed in \cite{Beccaria:2022kxy}, for $Sp(N)$ gauge theories an alternative strategy is possible whenever the matrix model has only single-trace action, which is the case for the two theories in Table \ref{tab:MatterContent}. 
As we will review in Section \ref{sec:TodaChain}, this property ensures the existence of a Toda equation, which is a highly efficient computational tool both at finite $N$ and in the large-$N$ limit. Ultimately, this equation enables us to analytically compute several orders of the large-$N$ expansion for both \eqref{ISpN4} and \eqref{ISpN2}.

\section{The partition function and the Toda equation}
\label{sec:TodaChain}
It is well-known that the partition function of a unitary matrix model, whose potential consists solely of a linear combination of single-trace terms, satisfies recursive relations that define an integrable Toda-like hierarchy \cite{GERASIMOV1991565,Alvarez-Gaume:1991xsn,Morozov:2009uy}. Recently, these relations have been extended to symplectic matrix models \cite{Beccaria:2022kxy}, where they have been applied to compute the large-$N$ expansion of the free energy and the v.e.v. of the $\frac{1}{2}$-BPS Wilson loop \eqref{WilsonLoopSp}. Specifically, it was shown that the partition function $\mathcal{Z}_N$ of a theory with $Sp(N)$ gauge group and a single-trace potential satisfies the following Toda equation
\begin{align}
\partial_{y}^{2}\log \mathcal{Z}_N(y) = \frac{\mathcal{Z}_{N+1}(y)\mathcal{Z}_{N-1}(y)}{\mathcal{Z}_N(y)^2}\, \ ,
\label{TodaChain}
\end{align}
 with the boundary conditions
\begin{align}
\mathcal{Z}_{N=-1}(y) = 0\, , \qquad \mathcal{Z}_{N=0}(y)=1   
\end{align}
where
\begin{align}
y = \frac{(4\pi)^2N}{\lambda} \, \ .
\label{y}
\end{align}
From \eqref{TodaChain}, it easily follows that the free energy $F_N(y) = -\log \mathcal{Z}_{N}(y)$ satisfies the equation
\begin{align}
\partial^2_y F_N(y) = -\exp\Big[\!-F_{N+1}(y)+2F_N(y)-F_{N-1}(y)\Big]\, \ .
\label{TodaFreeEnergy}
\end{align}
Equations \eqref{TodaChain} and \eqref{TodaFreeEnergy} are very general and hold, in particular, for both theories in Table \ref{tab:MatterContent}. 
For $\mathcal{N}=4$ SYM, the solution of  equation \eqref{TodaFreeEnergy} reads \cite{Beccaria:2022kxy}
\begin{align}
F_N^{\mathcal{N}=4}(y) = \frac{N(2N+1)}{2}\log y - \log C_{N}^{\mathcal{N}=4}\, \ ,    \label{FN4}
\end{align}
with
\begin{align}
C_{N}^{\mathcal{N}=4}  = \frac{G(N+1)G(N+3/2)}{G\left({3}/{2}\right)}\, \ .      \end{align}

As we are going to see,
equation \eqref{TodaChain} allows to derive a Toda equation for many observables. In the following, we are considering two different cases. The first case involves the v.e.v. of operators. The second case concerns the computation of the functions \eqref{tfunctions}.

\subsection{Toda equations for v.e.v. of operators}
\label{Toda-chain for connected operators}
To discuss the Toda equation satisfied by the v.e.v.'s of operators, following \cite{Beccaria:2022kxy} we introduce the deformed partition function 
\begin{align}
\widehat{\mathcal{Z}}_{N}(\textbf{c};y) = {\mathcal C}_N \, \left(\frac{2}{y}\right)^{\frac{N(2N+1)}{2}}
\int da \exp\left[-\frac{1}{2}\sum_{n=1}^{\infty}c_{2n}\,
\left(\frac{2}{y}\right)^n\, \text{tr}a^{2n}\right]\, \ ,
\label{ZGeneralized}
\end{align}
where $\textbf{c}=\{c_2,c_4,\cdots\}$ are arbitrary parameters. This partition function still satisfies \eqref{TodaChain}. Fixing the parameters $c_{2n}$ to particular values $\widehat{c}_{2n}$, one can retrieve the partition function of the un-deformed theories. In particular, for $\cN=4$ SYM  one has
\begin{align}
    \widehat{c}_2=y \quad\mbox{and}\quad \widehat{c}_{2n}=0\quad\mbox{for}\quad n \geq 2
    \label{c2n4}
\end{align}
and for the $\cN=2$ model one has
\begin{align}
\widehat{c}_2=y \quad\mbox{and}\quad \widehat{c}_{2n}= 4 (-1)^n \,(2^{2n-2}-1)\frac{\zeta_{2n-1}}{n}\quad\mbox{for}\quad n \geq 2\,\  .
\label{c2n2}
\end{align}
Then, we consider a generic single-trace operator $A_N$ of the form
\begin{align}
    A_N(y)=-\frac{1}{2}\sum_{n=1}^\infty \left(\frac{2}{y}\right)^{n}\,x_{2n}\,\text{tr}a^{2n}
    \label{AN}
\end{align}
where $x_{2n}$ are arbitrary coefficients. It is straightforward to realize that the v.e.v. of $A_N$ in the un-deformed theories is given by
\begin{align}
  \mathcal{A}_N(y)\,\equiv\,  \big\langle A_N\big\rangle = \partial_{X} \log \widehat{\mathcal{Z}}_{N}(\textbf{c};y)\,\Big|_{{c_{2n}} = \widehat c_{2n}}
  \label{vevAN}
\end{align}
where
\begin{align}
\partial_{X} = \sum_{n=1}^\infty x_{2n} \, \frac{\partial}{\partial {c_{2n}}} \, \ .
\label{dx}
\end{align}
By applying this differential operator to the equation \eqref{TodaChain} generalized to the deformed partition functions, it is easy to 
show that the v.e.v. $\mathcal{A}_N(y)$ satisfies a Toda-like equation, namely:
\begin{align}
\partial_{y}^2 \, {\mathcal A}_N(y) = - \partial_{y}^2 F_{N}(y)\,\Big[{\mathcal A}_{N+1}(y) - 2{\mathcal A}_N(y) +
{\mathcal A}_{N-1}(y)\Big]  \ .
\label{ATodaGeneral}
\end{align}

Also the case of double-trace operators can be considered in a similar way by simply further differentiating \eqref{ATodaGeneral}. In this way it is possible to obtain other Toda relations for the connected v.e.v.'s of products of two operators $A_{N}$ and $B_{N}$ of the form \eqref{AN} (whose v.e.v's are ${\mathcal A}_N$ and ${\mathcal B}_N$). Indeed, one finds \footnote{Here for simplicity we have not explicitly written the dependence on $y$.}
\begin{equation}
\partial_y^2 \mathcal{D}_N = -\partial_y^2 F_N\,\Big[\mathcal{D}_{N+1} -2\mathcal{D}_N +\mathcal{D}_{N-1}+({\mathcal A}_{N+1} -2{\mathcal A}_N +{\mathcal A}_{N-1})
({\mathcal B}_{N+1} - 2{\mathcal B}_N +{\mathcal B}_{N-1})\Big]\, \ ,
\label{TodaAB}
\end{equation}
where we have defined
\begin{align}
\mathcal{D}_N \equiv \langle A_N\, B_N \rangle_{\rm con} =
\langle A_N\, B_N \rangle - 
\mathcal{A}_N\, \mathcal{B}_N\, \ .
\label{Kcon}    
\end{align}

As already mentioned, equation \eqref{TodaChain}, and consequently equations \eqref{ATodaGeneral} and \eqref{TodaAB}, hold for both $\mathcal{N}=4$ SYM and the $\mathcal{N}=2$ model. However, to illustrate how to solve these equations in the large-$N$ limit, we focus first on the $\mathcal{N}=4$ SYM theory which is simpler, and denote the corresponding
quantities by a subscript $_0$. Using the $\mathcal{N}=4$ free energy given in \eqref{FN4}, we can rewrite equation \eqref{ATodaGeneral} as
\begin{align}
y^2 \partial_{y}^2 \, {\mathcal A}_{0;N}(y) = - 
\frac{N(2N+1)}{2}\,
\Big[{\mathcal A}_{0;N+1}(y) - 2{\mathcal A}_{0;N}(y) +{\mathcal A}_{0;N-1}(y)\Big]  \, \ .
\label{AN4withy}
\end{align}
In the large $N$-limit it's useful to express the above relation in terms of the 't Hooft coupling $\lambda$ of the $Sp(N)$ theory given in \eqref{y}, obtaining
\begin{align}
\big(\lambda^2\partial_{\lambda}^{2}+2\lambda\partial_{\lambda}\big) {\mathcal A}_{0;N} (\lambda)  = 
\frac{N(2N+1)}{2} \bigg[{\mathcal A}_{0;N+1} \Big(\lambda\frac{N+1}{N}\Big)-2\,{\mathcal A}_{0;N} (\lambda)+{\mathcal A}_{0;N-1} \Big(\lambda\frac{N-1}{N}\Big)\bigg]\, .
\label{TodaAN4}
\end{align}
In the same way the Toda equation for $\mathcal{D}_{0;N}$ becomes 
\begin{align}
\big(\lambda^2\partial_{\lambda}^2+2\lambda\partial_{\lambda}\big) \mathcal{D}_{0;N} (\lambda)  &=
\frac{N(2N+1)}{2}\,\bigg[\mathcal{D}_{0;N+1}\Big(\lambda\frac{N+1}{N}\Big)-2\,\mathcal{D}_{0;N}(\lambda)+\mathcal{D}_{0;N-1}\Big(\lambda\frac{N-1}{N}\Big)\bigg] \nonumber \\
& ~~
+
\frac{2}{N(2N+1)} \,\Big[\big(\lambda^2\partial_{\lambda}^2+2\lambda\partial_{\lambda}\big){\mathcal A}_{0;N} (\lambda)\Big]\,\Big[\big(\lambda^2\partial_{\lambda}^2
+2\lambda\partial_{\lambda}\big){\mathcal B}_{0;N}(\lambda)\Big]\, \ .
\label{TodaABN4}
\end{align}
To explicitly solve the equations \eqref{TodaAN4} and \eqref{TodaABN4} in the large $N$-limit, we make an Ansatz and fix the boundary conditions at $\lambda=0$
according to the operator we are considering. 
Let us now discuss some specific examples. 

The first case is that of the circular Wilson loop (\ref{WilsonLoopSp}), which was analyzed in detail in \cite{Beccaria:2022kxy} and is particularly interesting for the evaluation of the integrated correlator \eqref{ISpN4}. This operator is not of the form (\ref{AN}) since it contains contributions from the trace of the identity ($\text{tr}a^0=2N$) and from the traces of odd powers of $a$. However, taking into account that the latter have vanishing expectation values, it is immediate to see that 
the v.e.v. of $W_N$, namely $\mathcal{W}_{0;N}(y)$, is such that 
\begin{align}
{\mathcal W}_{0;N}(y) -2N= \partial_{X} \log \widehat{\mathcal{Z}}_{N}(\textbf{c};y)\,\Big|_{c_{2n} = \widehat c_{2n}}
\label{wv}
\end{align}
where 
\begin{align}
\partial_{X} = - 2\sum_{n=1}^{\infty}\frac{(2 \pi)^{2n}}{(2n)!} \, \frac{\partial}{\partial {c_{2n}}} \,
\end{align}
and $\widehat{c}_{2n}$ are as in \eqref{c2n4}. Thus, ${\mathcal W}_{0;N}$
satisfies the Toda equation \eqref{TodaAN4}.

The exact results for the leading term in the large-$N$ expansion of ${\mathcal W}_{0;N}$ given in \cite{Fiol_2014,Erickson_2000}, suggest to introduce the following Ansatz
\begin{align}
\mathcal W_{0;N} (\lambda) = N\,\mathcal{W}^{(0)}_0(\lambda) + \mathcal{W}^{(1)}_0(\lambda) + \frac{\mathcal{W}^{(2)}_0(\lambda)}{N} + O\Big(\frac{1}{N^2}\Big)\, \ 
\label{WLargeN}
\end{align}
where \cite{Fiol_2014,Erickson_2000}
\begin{align}
\mathcal{W}^{(0)}_0(\lambda) = \, \frac{4}{\sqrt{\lambda}}I_1(\sqrt{\lambda})\, \ ,
\label{w0}
\end{align}
with $I_1(\sqrt{\lambda})$ being a modified Bessel functions of the first kind, and to impose the boundary conditions
\begin{align}
\mathcal{W}^{(j)}_0(0) = 0
\end{align}
for $j=1,2,\ldots$
Exploiting the Toda equation \eqref{TodaAN4}, all sub-leading terms in \eqref{WLargeN} can be obtained from  $\mathcal{W}^{(0)}_0$. For instance, the first two sub-leading terms are
\begin{subequations}
\begin{align}
\mathcal{W}^{(1)}_0(\lambda) \,=\, & \frac{1}{4}\big(1+\lambda\partial_{\lambda}\big)\mathcal{W}^{(0)}_0(\lambda) -\frac{1}{2}\, \ , \\[0.5em]
\mathcal{W}^{(2)}_0(\lambda) \,=\, & \frac{1}{48}\big(3\,\lambda^2\partial_{\lambda}^2+\lambda^3\partial_{\lambda}^3\big)\mathcal{W}^{(0)}_0(\lambda)\, \ .
\end{align}
\label{W12}%
\end{subequations}

The same pattern holds for all operators whose v.e.v. satisfies equation \eqref{TodaAN4}. For our future purposes, it is useful to study
the operator ${M}_N$ defined in \eqref{Msp}.
Note that this operator is not the one that describes the quadratic mass-deformation of the $\mathcal{N}=2^{\star}$ theory, but its v.e.v. in $\mathcal{N}=4$, which we denote as 
$\mathcal{M}_{0;N}$, will be instrumental to compute the integrated correlator \eqref{ISpN2} in the $\mathcal{N}=2$ model as we will discuss in Section \ref{sec:I4dN2}. 
The operator ${M}_N$ has the form \eqref{AN}; thus its v.e.v. $\mathcal{M}_{0;N}$ can be obtained as in \eqref{vevAN} with
\begin{align}
\partial_{X} = 2\sum_{n=1}^{\infty}(-1)^n(2n+1)\zeta_{2n+1}
\frac{\partial}{\partial c_{2n}}\,\, .
\label{dxw}
\end{align}
and consequently it satisfies the Toda equation \eqref{TodaAN4}. In analogy with the previous case, we assume that the large-$N$ expansion of $\mathcal{M}_{0;N}(\lambda)$ takes the following form
\begin{align}
\mathcal{M}_{0;N}(\lambda) = N\,\mathcal{M}^{(0)}_0(\lambda) + \mathcal{M}^{(1)}_0(\lambda) + \frac{\mathcal{M}^{(2)}_0(\lambda)}{N} + O\Big(\frac{1}{N^2}\Big)\, \ 
\label{MLargeN}
\end{align}
supplemented with the boundary conditions
\begin{align}
\mathcal{M}^{(j)}_0(0) = 0\, , \qquad  \, j=0,1,2,\cdots  \, \ , 
\end{align}
which are dictated by the weak coupling expansion \eqref{Msp}. Inserting \eqref{MLargeN} into \eqref{TodaAN4},
all coefficients $\mathcal{M}^{(j)}_0$ for $j \geq 1$ can be written as functions of $\mathcal{M}^{(0)}_0$ and its derivatives. For example, for $j=1,2$, we get
\begin{subequations}
\begin{align}
\mathcal{M}^{(1)}_0(\lambda) \,=\, & \frac{1}{4}\big(1+\lambda\partial_{\lambda}\big)\mathcal{M}^{(0)}_0(\lambda)\, \ \label{M1 gauss} , \\[0.5em]
\mathcal{M}^{(2)}_0(\lambda) \,=\,& \frac{1}{48}\big(3\lambda^2\partial_{\lambda}^2+\lambda^3\partial_{\lambda}^3\big)\mathcal{M}^{(0)}_0(\lambda)\, \ . \label{M2 gauss}    
\end{align}
\label{M12}%
\end{subequations}
Notice that, exactly as for the Wilson loop, the leading coefficient $\mathcal{M}^{(0)}_0$ is not fixed by the Toda equation. However, it can be determined from \eqref{Msp} by using the large-$N$ expansion of the v.e.v. of $\text{tr}a^{2k}$ given in \cite{Billo:2024ftq}, namely
\begin{align}
t_{2k} = \langle \text{tr}a^{2k} \rangle_0 = \frac{N^{k+1}}{2^{k-1}}\frac{(2k)!}{k!(k+1)!} + O(N^{k})\, .    \label{t2k}
\end{align}
After some steps, described in Appendix \ref{App:Details on N=4 SYM-like theory}, we obtain
\begin{align}
\mathcal{M}^{(0)}_0(\lambda) = - 2\int_0^{\infty}dt\,\frac{\text{e}^t\,t}{(\text{e}^t-1)^2}\bigg[\frac{4\pi}{\sqrt{\lambda}\,t}J_1\bigg(\frac{\sqrt{\lambda}\,t}{2\pi}\bigg)-1\bigg]\, \ .
\label{M0}
\end{align}

The last case we consider is that of the double-trace operator $W_NM_N$,
whose connected v.e.v. in $\mathcal{N}=4$ (similar to the one appearing in the numerator of \eqref{ISpN4}), 
is given by:
\begin{align}
\mathcal{K}_{0;N}(\lambda)\,\equiv\,\big\langle  W_N\,  M_N \big\rangle_0 - \mathcal{W}_{0;N}(\lambda)\,\mathcal{M}_{0;N}(\lambda)\,\ .
\label{k0n}
\end{align}
This function satisfies the Toda equation \eqref{TodaABN4} which, in the large-$N$ limit, can be solved as discussed above.
Since at leading order $\mathcal{K}_{0;N}(\lambda)$ is $O(N^0)$, we can write
\begin{align}
\mathcal{K}_{0;N}(\lambda) = \mathcal{K}^{(0)}_0(\lambda) + \frac{\mathcal{K}^{(1)}_0(\lambda)}{N}+ \frac{\mathcal{K}^{(2)}_0(\lambda)}{N^2} + O\Big(\frac{1}{N^3}\Big)\, \
\label{Ansatz K},   
\end{align}
with the boundary conditions
\begin{align}
\mathcal{K}^{(i)}_0(0) = 0 \qquad \text{for} \qquad i \geq 0\, \ .
\label{Kboundary}
\end{align}
Substituting \eqref{Ansatz K} into \eqref{TodaABN4}, after some simplifications we obtain 
\begin{align}
\partial_{\lambda}\mathcal{K}^{(0)}_0(\lambda) = \frac{1}{2 \lambda}\, 
\Big[\big(\lambda^2\partial_{\lambda}^2+2\lambda\partial_{\lambda}\big){\mathcal W}^{(0)}_0 (\lambda)\Big]  
\Big[\big(\lambda^2\partial_{\lambda}^2+2\lambda\partial_{\lambda}\big){\mathcal M}^{(0)}_0 (\lambda)\Big]\;.
\end{align}
Using \eqref{Kboundary} together with \eqref{w0} and \eqref{M0} we find
\begin{align}
\mathcal{K}^{(0)}_0(\lambda) = \int_0^{\infty}dt\, \frac{\text{e}^t\,t}{(1-\text{e}^t)^2}\,\frac{t\sqrt{\lambda}}{4\pi^2+t^2}\bigg[2\pi\,I_2(\sqrt{\lambda})\,J_1\bigg(\frac{t\sqrt{\lambda}}{2\pi}\bigg)+t\,I_1(\sqrt{\lambda})J_2\bigg(\frac{t\sqrt{\lambda}}{2\pi}\bigg)\bigg]\, \ . 
\label{K0}
\end{align}
All other coefficients $\mathcal{K}^{(i)}_0$ with $i \geq 1$ are fixed by \eqref{TodaABN4} to be functions only of $\mathcal{M}^{(0)}_0$ and $\mathcal{W}^{(0)}_0$ and their derivatives. For example,
\begin{align}
\mathcal{K}^{(1)}_0(\lambda) \,&= \,  \frac{\lambda^2}{2}\partial_{\lambda}\mathcal{M}^{(0)}(\lambda)\partial_{\lambda}\mathcal{W}^{(0)}_0(\lambda) + \frac{\lambda^3}{4}\Big(\partial_{\lambda}\mathcal{M}^{(0)}_0(\lambda)\partial_{\lambda}^2\mathcal{W}^{(0)}_0(\lambda)+\partial_{\lambda}^2\mathcal{M}^{(0)}_0(\lambda)\partial_{\lambda}\mathcal{W}^{(0)}_0(\lambda)\Big) \nonumber \\
& + \frac{\lambda^4}{8}\partial_{\lambda}^2\mathcal{M}^{(0)}_0(\lambda)\partial_{\lambda}^2\mathcal{W}^{(0)}_0(\lambda)\, \ .        
\end{align}
We computed these coefficients up to $\mathcal{K}_0^{(10)}$, which corresponds to the order $O(N^{-10})$ in the large-$N$ limit. 
The explicit results are provided in an ancillary Mathematica file.

It is crucial to note that this method of solving the Toda equations in the large $N$-limit crucially relies on the fact that all v.e.v.'s are connected. If instead one needs to consider non-connected v.e.v.'s, the corresponding Toda equation becomes more involved and the above procedure does not work. For this reason, in 
Section \ref{sec:IN4} we propose a different strategy to deal with such cases.

\subsection{Toda equation for the v.e.v.'s of multiple traces}
\label{subsec:Todafort}
The method outlined in Section \ref{Toda-chain for connected operators} can be used also to compute the connected v.e.v.'s of the multiple traces (\ref{tfunctions}). It can be easily shown that they satisfy the Toda relations \eqref{ATodaGeneral} and \eqref{TodaAB}. In fact, by using the definition \eqref{tfunctions} and the generalized partition function \eqref{ZGeneralized}, the $\mathcal{N}=4$ v.e.v.'s of single and connected double traces at finite $N$ can be written respectively as
\begin{subequations}
\begin{align}
\Big(\frac{2}{y}\Big)^k\, t_{2k}^{(N)} &= -2\frac{\partial}{\partial c_{2k}}\log \mathcal{\widehat{Z}}_N(\textbf{c};y)\,\Big|_{c_{2n} = \widehat c_{2n}}\, \ ,
\label{Deft2k}\\[0.5em]
\Big(\frac{2}{y}\Big)^{k_1+k_2} t^{(N)\,c}_{2k_1,2k_2} &\equiv \Big(\frac{2}{y}\Big)^{k_1+k_2} \Big(t_{2k_1,2k_2}^{(N)} - t_{2k_1}^{(N)}t_{2k_2}^{(N)}\Big) = 4\, \frac{\partial^2}{\partial c_{2k_1}\partial c_{2k_2}}\log \mathcal{\widehat{Z}}_N(\textbf{c};y)\,\,\Big|_{c_{2n} = \widehat c_{2n}} \, ,
\label{Deft2k1k2Connected}
\end{align}%
\end{subequations}
where $\widehat{c}_{2n}$ are as in (\ref{c2n4}).
After working out the $y$ dependence in corresponding Toda equations, we obtain
\begin{align}
 k(k+1)t^{(N)}_{2k} &=
\frac{N(2N+1)}{2}\, \left[t^{(N+1)}_{2k}-2\,t^{(N)}_{2k}+t^{(N-1)}_{2k}\right]\, \ ,
 \label{TodaTracesN41pt} \\[0.5em]
(k_1+k_2+1)(k_1+k_2)t_{2k_1,2k_2}^{(N)\,c} &=
\,\frac{N(2N+1)}{2}\,\left[t_{2k_1,2k_2}^{(N+1)\,c}-2\,t_{2k_1,2k_2}^{(N)\,c}+t_{2k_1,2k_2}^{(N-1)\,c}\right] \nonumber\\
 &\quad+
 \frac{2}{N(2N+1)}\,\prod_{i=1}^{2}k_i(k_i+1)t_{2k_i}^{(N)} \, \ .
 \label{eqToda2ptcon} 
\end{align}
Following the same procedure, the Toda equations for higher point connected functions can also be derived. In general, it can be shown that the equation for the connected $t^{(N)\,c}_{2k_1,\cdots,2k_p}$ function involves all the lower connected correlators. For this reason we begin by looking for a solution of the equation \eqref{TodaTracesN41pt} and, in the large-$N$ limit, we make the following Ansatz
\begin{align}
t_{2k}^{(N)} = \sum_{q=1}^{k+1}d_{\text{1pt};q}(k)\,N^{k+2-q}\, \ .
\label{1ptLargeNAnsatz}
\end{align}
By comparing with \eqref{t2k}, we see that the leading coefficient is given by
\begin{align}
d_{\text{1pt};1}(k) = \frac{2^{1-k} (2 k)!}{k!\, (k+1)!}\, \ ,
\label{d1pt}
\end{align}
while the sub-leading coefficients $d_{\text{1pt};q}$ for $q \geq 2$ are determined by requiring that the Ansatz \eqref{1ptLargeNAnsatz} satisfies equation \eqref{TodaTracesN41pt}. For example, for the first values of $q$ we obtain
\begin{align}
d_{\text{1pt};2}(k) = \frac{(k+1)}{4}d_{\text{1pt};1}(k)\, , \qquad  d_{\text{1pt};3}(k) = \frac{(k^2-1) k}{48}d_{\text{1pt};1}(k) \, \ ,
\end{align}
We are now ready to consider equation \eqref{eqToda2ptcon}.  By using the
large-$N$ Ansatz
\begin{align}
t_{2k_1,2k_2}^{(N)\,c} = \sum_{q=1}^{k_1+k_2}d_{\text{2pt};q}(k_1,k_2)N^{k_1+k_2+1-q}\, \ , 
\label{t2cAnsatz}
\end{align}
we find that the coefficients $d_{\text{2pt};q}(k_1,k_2)$ are of the form
\begin{align}
d_{\text{2pt};q}(k_1,k_2)=\left[\prod_{i=1}^{2}2k_i(2+2k_i)\,d_{\text{1pt};1}(k_i)\right]\widetilde{d}_{\text{2pt};q}(2k_1,2k_2)\, \ ,    
\end{align}
where $\widetilde{d}_{\text{2pt};q}$ can be determined iteratively for any $q$. For example, for the first values of $q$, we obtain
\begin{align}
& \widetilde{d}_{\text{2pt};1}(k_1,k_2)=\frac{1}{16(k_1+k_2)}, \qquad \widetilde{d}_{\text{2pt};2}(k_1,k_2)=\frac{1}{128}, \nonumber \\[0.5em]
& \widetilde{d}_{\text{2pt};3}(k_1,k_2) = \frac{k_1^2+k_2 k_1-4 k_1+k_2^2-4 k_2+4}{6144}\,  .    
\end{align}
Once the large-$N$ expansions of the v.e.v. of traces are known, they can be used, by employing \eqref{PK}, to  determine the corresponding expansions for the correlation functions among the $\mathcal{P}$ operators,  which will be widely used in Section  \ref{sec:IN4}. For example, using the expression \eqref{t2cAnsatz}, we have
\begin{align}
&\langle \mathcal{P}_{2n_1}\,\mathcal{P}_{2n_2} \rangle_0 =  \nonumber\\&64\sqrt{n_1\,n_2}\left[\prod_{i=1}^{2}\sum_{\ell_i=0}^{n_i-1}\frac{(-1)^{\ell_i}(2n_i-\ell_i-1)!}{(\ell_i)!(n_i-\ell_i)!(n_i-\ell_i-1)!}\right]\sum_{q=1}^{n_1-\ell_1+n_2-\ell_2}\widetilde{d}_{\text{2pt};q}(2n_1-2\ell_1,2n_2-2\ell_2)N^{1-q}\, \ .    
\end{align}
It is worth noting that, thanks to the explicit expression of the coefficients $\widetilde{d}_{\text{2pt};q}(k_1,k_2)$, the sums over $\ell_1$ and $\ell_2$ can be computed analytically and, apart from the leading term, they reduce to evaluating finite sums of the following form
\begin{align}
f(n;p) = \sum_{\ell=0}^{n-1}\frac{(-1)^{\ell}(2n-\ell-1)!\, \ell^p}{\ell!\,(n-\ell)!\,(n-\ell-1)!} \quad \text{with} \quad p=0,1,2,\cdots\, \ .
\label{gMath}
\end{align}
For instance,  for the first values of $p$, we find
\begin{align}
f(n;0) = 1,\qquad f(n;1) = n-n^2, \qquad f(n;2)=\frac{1}{2} \left(n^4-4 n^3+3 n^2\right)\, \ .   
\end{align}
Using this technique, we computed the 2- and 3-point correlators among the $\mathcal{P}$ operators. Details on these computations and the first terms in their large-$N$ expansions are given in Appendix \ref{app:ConnectedCorrelators}, while the expansion coefficients up to order $O(N^{-8})$ can be found in the ancillary Mathematica file.

\section{The integrated correlator \texorpdfstring{$\mathcal{I}$}{}}
\label{sec:I4dN2}
In this Section we compute the large-$N$ expansion of the integrated correlator $\mathcal{I}$ defined in \eqref{ISpN2} for the $\mathcal{N}=2$ model. This analysis heavily relies on the technique based on Toda equations introduced in Section \ref{sec:TodaChain}. Specifically, we will show that the integrated correlator $\mathcal{I}$ is entirely determined by the solutions of the Toda equations for the free energy, the v.e.v.'s of the Wilson loop \eqref{WilsonLoopSp}, and the mass operator $M$ \eqref{Msp}. For this reason we start by deriving the large-$N$ expansions of these quantities.  

\subsection{Large-\texorpdfstring{$N$}{} computation using the Toda equations}
\label{subsec:TodaforN2}
As a first step we consider the free energy and, in order to make the paper self-contained, we review the large-$N$ analysis of \cite{Beccaria:2022kxy}, where the following Ansatz was proposed
\begin{align}
F_N(\lambda)&\, \equiv \,F_{N}^{\mathcal{N}=4}(\lambda)+ \Delta F_N(\lambda)\notag\\[1mm]
&= F_{N}^{\mathcal{N}=4}(\lambda) + NF^{(1)}(\lambda) +F^{(2)}(\lambda) + \frac{F^{(3)}(\lambda)}{N} + O\Big(\frac{1}{N^2}\Big)\, \ . 
\label{Fansatz}
\end{align}
Here $F_N(\lambda)$ denotes the free energy of the $\mathcal{N}=2$ model, while $F_{N}^{\mathcal{N}=4}(\lambda)$ is that of the $\mathcal{N}=4$ SYM theory given in \eqref{FN4}. 
The term $F^{(1)}$ is not fixed by requiring that the Ansatz \eqref{Fansatz}
satisfies the Toda equation \eqref{TodaFreeEnergy}.
However, since the free energy can be expressed in terms of the connected correlators of the interaction action \eqref{SOtilde} as
\begin{align}
\Delta F_N(\lambda) = \langle S_0 \rangle_0 - \frac{1}{2}\langle S_0\,S_0 \rangle_{0;\text{con}}  + \frac{1}{3!}  \langle S_0\,S_0\,S_0 \rangle_{0;\text{con}} + \cdots \;,
\end{align}
$F^{(1)}$ can be determined by   
\begin{align}
F^{(1)}(\lambda) &= \lim_{N \rightarrow\infty}\,\frac{1}{N}\,\Delta F_N(\lambda) = \lim_{N \rightarrow\infty}\,\frac{1}{N}\,\langle S_0 \rangle_{0} \notag\\[2mm]
&= \lim_{N \rightarrow \infty}\,\frac{1}{N} \bigg[ 4\sum_{k=1}^{\infty}(-1)^{k+1}\Big(\frac{\lambda}{8\pi^2N}\Big)^{k+1}(2^{2k}-1)\frac{\zeta_{2k+1}}{k+1}\,\langle\text{tr}\,a^{2k+2}\rangle_{0}\bigg]\, \ .   
\label{F1in}
\end{align}
From the large-$N$ expansion of $\langle \text{tr}\,a^{2k+2} \rangle_0$
given in \eqref{t2k}, after summing over $k$, we obtain
\begin{align}
F^{(1)}(\lambda) =  \frac{\log(2)}{2\pi^2}\,\lambda \,+ \, 4\int_{0}^{\infty}\frac{dt}{t}\frac{\text{e}^t}{(\text{e}^t+1)^2}\bigg[\bigg(\frac{2\pi}{\sqrt{\lambda}\,t}\bigg)\,J_1\bigg(\frac{\sqrt{\lambda}\,t}{\pi}\bigg)-1\bigg] \,  .
\label{F1}
\end{align}
The other coefficients $F^{(i)}$ with $i \geq 2$ are then determined by imposing the Toda equation \eqref{TodaFreeEnergy}. For example, the first two coefficients are 
\begin{subequations}
\begin{align}
\partial_{\lambda}F^{(2)}(\lambda ) & = -\frac{1}{4} \Big(\lambda \partial_{\lambda}^2F^{(1)}(\lambda )+2\partial_{\lambda}F^{(1)}(\lambda )\Big) \Big(\lambda ^2\partial_{\lambda}^2F^{(1)}(\lambda )+2 \lambda\partial_{\lambda}F^{(1)}(\lambda )-1\Big)\, ,\\[0.5em]
F^{(3)}(\lambda ) &= \frac{1}{48} \lambda ^2\bigg[\lambda \partial_{\lambda}^3F^{(1)}(\lambda )+3\partial_{\lambda}^2F^{(1)}(\lambda)\,+ \bigg. \nonumber \\
& \bigg. \quad \Big(2\lambda ^2\partial_{\lambda}^2F^{(1)}(\lambda )+4 \lambda\,\partial_{\lambda}F^{(1)}(\lambda)-3\Big) \Big(\lambda \partial_{\lambda}^2F^{(1)}(\lambda )+2\partial_{\lambda}F^{(1)}(\lambda )\Big)^2\bigg]\,  .
\end{align}
\label{F2andF3}%
\end{subequations}
We computed these coefficients up to $F^{(8)}$, corresponding to the $O(N^{-6})$ order. 

The next step is to find the large-$N$ expansion for the v.e.v. $\mathcal{M}(\lambda)$
of the operator $M$ defined in \eqref{Msp}. In analogy with the Gaussian case \eqref{MLargeN}, we make the Ansatz
\begin{align}
\mathcal{M}(\lambda) 
= N \mathcal{M}^{(0)}(\lambda) + \mathcal{M}^{(1)}(\lambda) + \frac{1}{N}\mathcal{M}^{(2)}(\lambda) + O\Big(\frac{1}{N^2}\Big)\, \; ,
\label{MLargeN2}
\end{align}
and then insert it into the Toda equation \eqref{ATodaGeneral} where we employ the previously determined expansion \eqref{Fansatz} of the free energy. Once again the leading coefficient $\mathcal{M}^{(0)}$ is not fixed by the Toda equation, but its explicit computation shows that it coincides with the Gaussian result \eqref{M0}. The remaining coefficients depend solely on $\mathcal{M}^{(0)}$, $F^{(1)}$ and their derivatives, and are fully determined after imposing the following boundary conditions at $\lambda = 0$
\begin{align}
\mathcal{M}^{(i)}(0)=0\, , \qquad i \geq 1 \quad .
\end{align}
For example for $i=1,2$ we find  
\begin{subequations}
\begin{align}
\partial_{\lambda}\mathcal{M}^{(1)}(\lambda) = &  -\frac{1}{4}\Big(2\lambda^2 \partial_{\lambda}^2F^{(1)}(\lambda )+4 \lambda\partial_{\lambda}F^{(1)}(\lambda )-1\Big) \Big(\lambda\partial_{\lambda}^2 \mathcal{M}^{(0)}+2\partial_{\lambda}\mathcal{M}^{(0)}\Big)\, \ , 
\label{M1int}\\[0.5em]
\mathcal{M}^{(2)}(\lambda) = & \, \frac{1}{48} \,\lambda^2\bigg[6\Big(\lambda \partial_{\lambda}^2F^{(1)}(\lambda )+2 \partial_{\lambda}F^{(1)}(\lambda )\Big)
\Big(\lambda^2\partial_{\lambda}^2F^{(1)}(\lambda )+2\lambda\partial_{\lambda}F^{(1)}(\lambda )-1\Big) \nonumber \bigg. \\
& \bigg.\Big(\lambda \partial_{\lambda}^2\mathcal{M}^{(0)}+2 \partial_{\lambda}\mathcal{M}^{(0)}\Big)  
+\lambda \partial_{\lambda}^3\mathcal{M}^{(0)}+3 \partial_{\lambda}^2\mathcal{M}^{(0)}\bigg]\, \ .
\label{M2int}
\end{align}
\label{M1andM2interacting}%
\end{subequations}
Comparing these results with their Gaussian counterparts \eqref{M12}, we note that they are more complicated because of the $\lambda$-dependence of $F^{(1)}$.

Next we consider the v.e.v. ${\mathcal W}_N(\lambda)$ of the circular Wilson loop whose large-$N$ expansion is
\begin{align}
{\mathcal W}_N(\lambda) 
= N\,\mathcal{W}^{(0)}(\lambda) + \mathcal{W}^{(1)}(\lambda) + \frac{1}{N}\mathcal{W}^{(2)}(\lambda) + O\Big(\frac{1}{N^2}\Big)\, \; .
\label{WLargeN2}    
\end{align}
Since ${\mathcal W}_N(\lambda)$ satisfies the same Toda equation as $\mathcal{M}_{N}(\lambda)$, the coefficients $\mathcal{W}^{(i)}$ for $i \geq 1$ can be obtained from \eqref{M1andM2interacting} with obvious replacements \footnote{In principle, both $\mathcal{M}^{(1)}$ and $\mathcal{W}^{(1)}$ are determined up to an integration constant. However, by imposing the boundary conditions satisfied by these functions at $\lambda=0$, one can demonstrate that these constants vanish.}. Also in this case, the leading term $\mathcal{W}^{(0)}$ coincides with its Gaussian counterpart \eqref{w0}. We were able to explicitly compute the coefficients 
$\mathcal{M}^{(i)}$ and $\mathcal{W}^{(i)}$ up to $i=6$. 

Finally, we derive the large-$N$ expansion of 
\begin{align}
\mathcal{K}_N (\lambda) \equiv \big\langle M_NW_N \big\rangle - \mathcal{M}_N(\lambda)\, \mathcal{W}_N(\lambda) \, \ .
\label{KN2}
\end{align}
In analogy with the Gaussian case 
\eqref{Ansatz K}, we make the Ansatz
\begin{align}
\mathcal{K}_N(\lambda) = \mathcal{K}^{(0)}(\lambda) + \frac{\mathcal{K}^{(1)}(\lambda)}{N}+ \frac{\mathcal{K}^{(2)}(\lambda)}{N^2} + O\Big(\frac{1}{N^3}\Big)\, \ ,  
\label{AnsatzK2}
\end{align}
and substitute it into the Toda equation \eqref{TodaAB}.
Imposing the boundary conditions 
\begin{align}
\mathcal{K}^{(i)}(0)=0\, , \qquad i \geq 0 \, ,
\end{align}
we can determine the coefficients $\mathcal{K}^{(i)}$ with $i \geq 0$. Again, we find that $\mathcal{K}^{(0)}$ coincides with the Gaussian result \eqref{K0}, while, for example, the next-to-planar term is
\begin{align}
\mathcal{K}^{(1)}(\lambda)= & -\frac{\lambda^2}{8}\Big(2\lambda^2 \partial_{\lambda}^2F^{(1)}(\lambda)+4 \lambda \partial_{\lambda}F^{(1)}(\lambda )-1\Big) \Big(\lambda \partial_{\lambda}^2\mathcal{M}^{(0)}(\lambda )+2 \partial_{\lambda}\mathcal{M}^{(0)}(\lambda )\Big)\nonumber\\
& \,\, \Big(\lambda  \partial_{\lambda}^2\mathcal{W}^{(0)}(\lambda )+2\partial_{\lambda}\mathcal{W}^{(0)}(\lambda)\Big) \;.
\label{K1interacting}
\end{align}
Proceeding in this way, we have computed the sub-leading coefficients in terms of $\mathcal{W}^{(0)}$, $\mathcal{M}^{(0)}$, $F^{(1)}$ and their derivatives up to $\mathcal{K}^{(6)}$.

We are now ready to consider the integrated correlator $\mathcal{I}(\lambda)$. 
A quick inspection of equations \eqref{WLargeN2} and \eqref{AnsatzK2} shows that the large-$N$ expansion of 
$\mathcal{I}(\lambda)$ must be of the form
\begin{align}
\mathcal{I}(\lambda) = \frac{1}{N}\,\mathcal{I}^{(0)}(\lambda) + \frac{1}{N^2}\,\mathcal{I}^{(1)}(\lambda) + \frac{1}{N^3}\,\mathcal{I}^{(2)}(\lambda)+ O\Big(\frac{1}{N^4}\Big)\, \ .
\label{IlargeN}    
\end{align}
By using \eqref{ISpN2}, it can be shown that the coefficients $\mathcal{I}^{(n)}$ can be determined iteratively through the relation
\begin{align}  
\mathcal{I}^{(n)}(\lambda)=  \frac{2\,\mathcal{K}^{(n)}(\lambda)}{ \mathcal{W}^{(0)}(\lambda) } -\sum_{i=1}^{n}\frac{ \mathcal{W}^{(i)} (\lambda)}{\mathcal{W}^{(0)}(\lambda)}\,\mathcal{I}^{(n-i)}(\lambda)\, .
\label{Icoefficients}
\end{align}
The coefficients $\mathcal{I}^{(n)}$ obtained in this way can be easily evaluated at weak coupling. For example, one finds
\begin{subequations}
\begin{align}
&\mathcal{I}^{(0)}(\lambda)  =  \frac{3 \zeta_3}{32 \pi ^2}\lambda^2- \frac{\lambda^3}{256 \pi ^4}\left(\pi ^2 \zeta_3+10 \zeta_5\right) +\frac{\lambda^4}{8192 \pi ^6}\left( 2 \pi ^4 \zeta_3+10 \pi ^2 \zeta_5+105 \zeta_7\right) + O(\lambda^5)\, \ , \\[0.5em]
& \mathcal{I}^{(1)}(\lambda) = \frac{3 \zeta_3}{64\pi ^2}\lambda^2 -\frac{3\lambda^3}{512 \pi ^4}\left(\pi ^2 \zeta_3+5 \zeta_5\right) + \frac{\lambda^4}{8192 \pi ^6}\left(5 \pi ^4 \zeta_3+20 \pi ^2 \zeta_5+105 \zeta_7-432 \zeta_3^2\right) + O(\lambda^5)\, \ .
\end{align}%
\end{subequations}
Equation \eqref{Icoefficients} can be used also to obtain $\mathcal{I}^{(n)}$ at strong-coupling, provided
the asymptotic expansions of $\partial_{\lambda}F_1$, $\partial_{\lambda}\mathcal{M}^{(0)}$, $\mathcal{W}^{(0)}$ and $\mathcal{K}^{(0)}$ are known.
In this regard, we notice that the strong-coupling expansion of $\mathcal{W}^{(0)}$ can be readily derived from \eqref{w0}, while those 
of $\partial_{\lambda}F_1$, $\partial_{\lambda}\mathcal{M}^{(0)}$ and $\mathcal{K}^{(0)}$ can be obtained from \eqref{F1}, \eqref{M0} and \eqref{K0} by applying the Mellin-Barnes technique discussed in Appendix \ref{SC}, yielding 
\begin{subequations}
\begin{align}
\partial_{\lambda}F^{(1)}(\lambda) &\underset{\lambda \rightarrow \infty}{\simeq}  \frac{\log(2)}{2\pi^2} - \frac{1}{2\,\lambda} +  \frac{\pi^2}{2\lambda^2}\, 
\;, \label{StrongMF1}\\[1mm]
\partial_{\lambda}\mathcal{M}^{(0)}(\lambda) &\underset{\lambda \rightarrow \infty}{\simeq}\, \frac{1}{\lambda} - \frac{4\pi^2}{3\lambda^2}\,\;,\\[2mm]
\mathcal{K}^{(0)}(\lambda) &\underset{\lambda \rightarrow \infty}{\simeq} I_0(\sqrt{\lambda})\, \ .
\end{align}
\label{StrongM0F1}%
\end{subequations}
We note that the strong-coupling expansions of $\partial_{\lambda}F^{(1)}$ and $\partial_{\lambda}\mathcal{M}^{(0)}$ contain only a finite number of terms.
This is to be contrasted with what happens for a generic observable for which the strong-coupling expansion is in general an asymptotic series
in $1/\lambda$, see e.g. \cite{Beccaria:2022ypy}.
Using \eqref{StrongM0F1}, the large-$\lambda$ expansions of all coefficients $\mathcal{K}^{(i)}$ and $\mathcal{W}^{(i)}$ can be determined via the corresponding Toda equations, and the strong-coupling expansion of  $\mathcal{I}^{(n)}$ can be subsequently obtained.
We have explicitly performed these calculations up to  $\mathcal{I}^{(6)}$.
Although conceptually straightforward, we point out that this procedure has a very high computational cost, which forces us to stop at the order $O(N^{-7})$. In Section \ref{sec:strong2}, we outline a different and more efficient method to generate the large-$N$ expansion of the coefficients $\mathcal{I}^{(n)}$.

\subsection{An alternative route to strong coupling}
\label{sec:strong2}
A computationally more efficient strategy to obtain the strong-coupling expansions of the coefficients $\mathcal{I}^{(n)}$ is based on the analysis performed in \cite{Beccaria:2022kxy}. To describe it, we first observe that the large-$N$ expansion of the free energy at strong coupling can be resummed in a closed form \cite{Beccaria:2021ism} and that
\begin{align}
\partial_y^2F_N(y) = - \frac{\left(  N+ \frac{1}{2}  \right)^2-\frac{1}{16}}{(y+8\log2)^2} \, \ .
\label{devFStrong}
\end{align}
Expanding in inverse powers of $N$ and $\lambda$, one can obtain from this equation the strong-coupling values of the coefficients $F^{(i)}$. In particular one can show that 
$\partial_{\lambda}F^{(1)}$ is given by \eqref{StrongMF1}.
Quite remarkably, if in \eqref{devFStrong} one performs the redefinitions
\begin{align}
N+ \frac{1}{4}= N'\, , \qquad   y+8\log2 = y' \;,
\label{rescalingNy}
\end{align}
one can check that $F_N(y)$ satisfies the defining condition for the free energy of $\mathcal{N}=4$ SYM that is obtained by substituting \eqref{FN4} in \eqref{TodaFreeEnergy}.
Once expressed in terms of $\lambda$, the redefinitions \eqref{rescalingNy} correspond to introducing an effective coupling constant
\begin{align}
   \lambda' =\frac{\lambda}{1+\frac{\log(2)\lambda}{2\pi^2N}}\left(\frac{N+\frac{1}{4}}{N}\right)\, \ .
   \label{rescalinglambda}
\end{align}
It is easy to see that this property holds in all Toda equations of the interacting theory. This implies that the strong-coupling solutions 
of the Toda equations of the $\mathcal{N}=2$ model can be obtained from those of a $\mathcal{N}=4$ SYM theory with group $Sp(N')$ and 't Hooft coupling $\lambda'$.

We are therefore led to the following procedure. First, we define
\begin{align}
\mathcal{I}_0(\lambda') =2 \, \frac{\mathcal{K}_{0,N'}(\lambda')}{\mathcal{W}_{0,N'}(\lambda')}\, \  \label{I0new}   
\end{align}
which is the analogue of the integrated correlator \eqref{ISpN2}, with the key difference that the expectation values are computed in a $Sp(N')$ Gaussian matrix model 
with coupling $\lambda'$. Then, we consider the expansion
\begin{align}
\mathcal{I}_0(\lambda') = \frac{1}{N'}\,\mathcal{I}_0^{(0)}(\lambda') + \frac{1}{N'^{\,2}}\,\mathcal{I}^{(1)}(\lambda') + \frac{1}{N'^{\,3}}\,\mathcal{I}^{(2)}(\lambda')+ 
O\Big(\frac{1}{N'^{\,4}}\Big)\, \ .
\label{IlargeN'}    
\end{align}
where the coefficients $\mathcal{I}^{(n)}_0(\lambda')$ can be computed from a recursion relation analogous to \eqref{Icoefficients}, namely
\begin{align}
\mathcal{I}^{(n)}_{0}(\lambda')=  \frac{2\,\mathcal{K}^{(n)}_0(\lambda')}{ \mathcal{W}^{(0)}_0 (\lambda')} -\sum_{i=1}^{n}\frac{ \mathcal{W}^{(i)}_0(\lambda') }{\mathcal{W}^{(0)}_0(\lambda')}\,\,\mathcal{I}^{(n-i)}_{0}(\lambda')\, .
\label{I0coefficients}
\end{align}
Using the results for $\mathcal{K}^{(n)}_0$ and $\mathcal{W}^{(i)}_0 $ given in Section \ref{Toda-chain for connected operators}, we easily 
obtain the following large-$\lambda'$ expansions  
\begin{subequations}
\begin{align}
& \mathcal{I}^{(0)}_{0} \underset{\lambda' \rightarrow \infty}{\simeq}\, \frac{\lambda'^{1/2}}{2}+\frac{1}{4}+\frac{3}{16\lambda'^{1/2}}+\frac{3}{16\lambda'} + O(\lambda'^{-3/2})\, \ ,
\label{I0N4}\\[0.5em]
& \mathcal{I}^{(1)}_{0} \underset{\lambda' \rightarrow \infty}{\simeq}\, -\frac{\lambda'^{1/2}}{16}-\frac{1}{16}-\frac{9}{128\lambda'^{1/2}}-\frac{3}{32\lambda'} +  O(\lambda'^{-3/2})\, \ , \\[0.5em]
&  \mathcal{I}^{(2)}_{0} \underset{\lambda' \rightarrow \infty}{\simeq}\, -\frac{\lambda'^{3/2}}{768}+\frac{3\lambda'}{128}+\frac{21\lambda'^{1/2}}{2018}+\frac{7}{512}+\frac{615}{32768\lambda'^{1/2}} +O(\lambda'^{-1})\, \ , \\[0.5em]  
& \mathcal{I}^{(3)}_{0} \underset{\lambda' \rightarrow \infty}{\simeq}\, \frac{\lambda'^{3/2}}{2048}-\frac{3\lambda'}{512}-\frac{25\lambda'^{1/2}}{16384}-\frac{5}{2048}-\frac{945}{262144\lambda'^{1/2}} + O(\lambda'^{-1})\, \ , \\[0.5em]
& \mathcal{I}^{(4)}_{0} \underset{\lambda' \rightarrow \infty}{\simeq}\, \frac{3\lambda'^{5/2}}{327680}-\frac{3\lambda'^2}{16384}+\frac{1475\lambda'^{3/2}}{1572864}+\frac{9\lambda'}{32768} +O(\lambda'^{1/2})\, \ , \\[0.5em]
&  \mathcal{I}^{(5)}_{0} \underset{\lambda' \rightarrow \infty}{\simeq}\, -\frac{3\lambda'^{5/2}}{524288}+\frac{9\lambda'^2}{65536}-\frac{3815 \lambda'^{3/2}}{4194304}+\frac{39\lambda'}{32768} + O(\lambda'^{1/2}) \\[0.5em]  
&  \mathcal{I}^{(6)}_{0} \underset{\lambda' \rightarrow \infty}{\simeq}\, -\frac{5\lambda'^{7/2}}{58720256}+\frac{3\lambda'^3}{1048576}-\frac{2191\lambda'^{5/2}}{67108864}+\frac{75\lambda'^2}{524288} +O(\lambda'^{3/2})\, \ \\[0.5em]
&  \mathcal{I}^{(7)}_{0} \underset{\lambda' \rightarrow \infty}{\simeq}\, \frac{5\lambda'^{7/2}}{67108864} -\frac{3\lambda'^3}{1048576} + \frac{20727 \lambda'^{5/2}}{536870912} -\frac{495 \lambda'^2}{2097152} + O(\lambda'^{3/2})\, \ , \\[0.5em]
& \mathcal{I}^{(8)}_{0} \underset{\lambda' \rightarrow \infty}{\simeq}\, \frac{35\lambda'^{9/2}}{38654705664} -\frac{3\lambda'^4}{67108864} + \frac{28665 \lambda'^{7/2}}{34359738368}-\frac{2055 \lambda'^3}{268435456} + O(\lambda'^{5/2})\, \ , \\[0.5em]
& \mathcal{I}^{(9)}_{0} \underset{\lambda' \rightarrow \infty}{\simeq}\, -\frac{35\lambda'^{9/2}}{34359738368} +\frac{15\lambda'^4}{268435456} + O(\lambda'^{7/2})\, \ .
\end{align}
\label{ICoefficientsStromgN4like}%
\end{subequations}
Although it is not our final goal, it is interesting also to analyze the properties of these expansions. 
In particular, using the results of \cite{Martin_2010}, we observe that the leading term \eqref{I0N4} can be expressed compactly as
\begin{align}
\mathcal{I}^{(0)}_{0}(\lambda') \underset{\lambda' \rightarrow \infty}{\simeq}\, \frac{\lambda'^{1/2}}{2}\sum_{n=0}^{\infty}\frac{c_n}{(4\lambda'^{1/2})^n}\,\ ,
\end{align}
with
\begin{align}
c_n =  \int_0^{\infty}dx\,\frac{2\,x^{n-2}}{K_1\left(\frac{x}{4}\right)^2+\pi^2I_1\left(\frac{x}{4}\right)^2}\, \
\label{cn_coefficients}
\end{align}
where $K_1(x)$ is the modified Bessel function of the second kind. As a consequence of the Toda equation, all remaining coefficients $\mathcal{I}^{(j)}_{0}$ for $j \geq 1$ are expressed in terms $\mathcal{I}^{(0)}_{0}$ and its derivatives. For instance, for $j=1,2$, we have
\begin{subequations}
\begin{align}
& \mathcal{I}^{(1)}_{0}(\lambda') \underset{\lambda' \rightarrow \infty}{\simeq}\, -\frac{1}{4}\left(1-\lambda'\partial_{\lambda'}\right)\mathcal{I}^{(0)}_{0}(\lambda') = -\frac{\lambda'^{1/2}}{16}\sum_{n=0}^{\infty}\frac{(1+n)c_n}{(4\lambda'^{1/2})^n}\, , \\
& \mathcal{I}^{(2)}_{0}(\lambda')\underset{\lambda' \rightarrow \infty}{\simeq}\, \frac{\lambda'}{384}+\left(\frac{1}{8}\mathcal{I}^{(0)}_{0}(\lambda') -\frac{\lambda'}{192}\right)\left(1-\lambda'\partial_{\lambda'}\right)\mathcal{I}^{(0)}_{0}(\lambda')\, \ .
\label{IhigherasIO}
\end{align}%
\end{subequations}
Finally, the strong-coupling limit of the large-$N$ expansion (\ref{IlargeN}) of the integrated correlator $\mathcal{I}$ of the $\mathcal{N}=2$ model follows from the large-$N'$ expansion (\ref{IlargeN'}) of $\mathcal{I}_0$, applying the rescalings \eqref{rescalingNy} and \eqref{rescalinglambda}, and re-expanding for $N \to \infty$.
In this way we find 
\begin{small}
\begin{subequations}
\begin{align}
& \mathcal{I}^{(0)} (\lambda)\underset{\lambda \rightarrow \infty}{\simeq}\,\frac{\lambda^{1/2}}{2}+\frac{1}{4}+\frac{3}{16\,\lambda^{1/2}}+\frac{3}{16\lambda} + O(\lambda^{-3/2})\, \ , \\[0.5em]
&  \mathcal{I}^{(1)}(\lambda) \underset{\lambda \rightarrow \infty}{\simeq}\, -\frac{\log (2)}{8\pi ^2}\lambda ^{3/2} +\left(\frac{3 \log (2)}{64 \pi^2}-\frac{1}{8}\right)\lambda^{1/2}+\frac{3 \log (2)}{32 \pi ^2}-\frac{1}{8}+ O(\lambda^{-1/2})\, \ ,\\[0.5em]
&  \mathcal{I}^{(2)} (\lambda)\underset{\lambda \rightarrow \infty}{\simeq}\, \frac{3\log ^2(2)}{64 \pi ^4}\lambda ^{5/2}-\left(\frac{3 \log^2(2)}{512 \pi ^4}-\frac{\log (2)}{32 \pi ^2}+\frac{1}{768}\right)\lambda ^{3/2}+\frac{3 \lambda }{256}+ O(\lambda^{1/2})\, \ ,\\[0.5em]
&  \mathcal{I}^{(3)}(\lambda) \underset{\lambda \rightarrow \infty}{\simeq}\, -\frac{5\log ^3(2)}{256 \pi^6}\lambda ^{7/2}+\left(\frac{3 \log ^3(2)}{2048 \pi ^6}-\frac{3 \log ^2(2)}{256 \pi ^4}+\frac{\log(2)}{1024 \pi ^2}\right) \lambda^{5/2} +O(\lambda^{3/2})\, \ , \\[0.5em]
&  \mathcal{I}^{(4)}(\lambda) \underset{\lambda \rightarrow \infty}{\simeq}\, \frac{35\log ^4(2)}{4096\pi^8}\lambda ^{9/2}- \left(\frac{15\log^4(2)}{32768\pi^8}-\frac{5 \log ^3(2)}{1024 \pi ^6}+\frac{5\log^2(2)}{8192\pi^4}\right)\lambda ^{7/2} + O(\lambda^{3})\, \ ,\\[0.5em]
& \mathcal{I}^{(5)}(\lambda) \underset{\lambda \rightarrow \infty}{\simeq}\, -\frac{63\log ^5(2)}{16384 \pi ^{10}}\lambda ^{11/2}+\left(\frac{21\log^5(2)}{131072 \pi^{10}}-\frac{35\log^4(2)}{16384\pi^8}+\frac{35\log^3(2)}{98304\pi^6}\right)\lambda^{9/2}+ O(\lambda^4)\, \ , \\[0.5em]
& \mathcal{I}^{(6)} (\lambda)\underset{\lambda \rightarrow \infty}{\simeq}\, \frac{231\log^6(2)}{131072\pi^{12}}\lambda^{13/2}-\left(\frac{63\log^6(2)}{1048576\pi^{12}}-\frac{63\log^5(2)}{65536\pi^{10}}+\frac{105\log^4(2)}{524288\pi^8}\right)\lambda^{11/2}+ O(\lambda^5)\, \ , \\[0.5em]
& \mathcal{I}^{(7)}(\lambda) \underset{\lambda \rightarrow \infty}{\simeq}\, -\frac{429 \log ^7(2)}{524288 \pi ^{14}}\lambda^{15/2} + \left(\frac{99 \log ^7(2)}{4194304 \pi
   ^{14}}-\frac{231 \log ^6(2)}{524288 \pi
   ^{12}}+\frac{231 \log ^5(2)}{2097152 \pi^{10}}\right)\lambda^{13/2} + O(\lambda^{6})\, \  , \\[0.5em]
& \mathcal{I}^{(8)}(\lambda) \underset{\lambda \rightarrow \infty}{\simeq} \frac{6435 \log ^8(2)}{16777216 \pi ^{16}}\lambda^{17/2} - \left(\frac{1287 \log ^8(2)}{134217728 \pi
   ^{16}}-\frac{429 \log ^7(2)}{2097152 \pi
   ^{14}}+\frac{1001 \log ^6(2)}{16777216 \pi
   ^{12}}\right)\lambda^{15/2}   +O(\lambda^{7})\, \ , \\[0.5em]
&  \mathcal{I}^{(9)}(\lambda) \underset{\lambda \rightarrow \infty}{\simeq} -\frac{12155 \log ^9(2)}{67108864 \pi ^{18}}\lambda^{19/2} +\left(\frac{2145 \log ^9(2)}{536870912 \pi
   ^{18}}-\frac{6435 \log ^8(2)}{67108864 \pi
   ^{16}}+\frac{2145 \log ^7(2)}{67108864 \pi
   ^{14}}\right)\lambda^{17/2} + O(\lambda^8)\, \ .
\end{align}
\label{IFINALExpansions}%
\end{subequations}
\end{small}%
This is our final result for the large-$N$ and strong coupling expansion of the integrated correlator \eqref{ISpN2}.
Of course, the first $\mathcal{I}^{(i)}$ coefficients coincide with the ones that have been obtained by directly solving the Toda equation
as described in Section \ref{subsec:TodaforN2}, but this second method based on an auxiliary $\mathcal{N}=4$ model is much more efficient. 
The same result can also be derived by performing the computation of the relevant vacuum expectation values 
using the basis of the $\mathcal{P}_{k}$ operators \eqref{PK} in the interacting theory. We performed this computation up to the order $O(N^{-4})$, finding perfect agreement with the expressions \eqref{IFINALExpansions}. We will give more details on this approach in the next Section, when discussing the integrated correlator of the $\mathcal{N}=4$ SYM theory.

\section{The integrated correlator \texorpdfstring{$\widetilde{\mathcal{I}}$}{}}
\label{sec:IN4}
We observed in Section \ref{sec:TodaChain} that the technique based on the Toda equations to compute the large-$N$ expansion 
of the integrated correlator 
is not directly applicable to the $\mathcal{N}=4$ SYM because of the presence of the double trace term in the mass operator \eqref{M1andM2}. In this Section, we describe a way to carry out this computation directly in
the massive matrix model of the $\mathcal{N}=2^*$ theory, building on the methods described in \cite{Billo:2024ftq}.
In particular we will show that the Toda relations \eqref{TodaTracesN41pt} and \eqref{eqToda2ptcon} 
allow us to  derive the large-$N$ expansion of the $\mathcal{N}=4$ integrated correlator $\widetilde{\mathcal{I}}(\lambda)$ at very high order.

\subsection{The large-\texorpdfstring{$N$}{} expansion of \texorpdfstring{$\widetilde{\mathcal{K}}$}{}}
Let us begin by considering the quantity
\begin{align}
\langle \widetilde{M}\,{W}_N \rangle_0 - \langle \widetilde{M} \rangle_0 \, \langle {W}_N \rangle_0  \,\equiv\, 
\widetilde{\mathcal{K}}_N(\lambda)
\end{align}
appearing in the right-hand side of \eqref{ISpN4}, which upon using \eqref{M1andM2} and \eqref{WilsonLoopSp} can be written as
\begin{align}
\widetilde{\mathcal{K}}_N(\lambda)=\sum_{i=1}^{2}  
\big\langle \widetilde{M}^{(i)}\, {W}_N \big\rangle_{0;\text{con}}\, \ , 
\label{StartN4}
\end{align}
where
\begin{subequations}
\begin{align}
\langle \widetilde{M}^{(1)}\,{W} \rangle_{0;\text{con}}   =&  -\frac{1}{2}\sum_{n=1}^{\infty}\sum_{k=0}^{\infty}\frac{1}{(2k)!}\Big(\frac{\lambda}{2N}\Big)^k\,(-1)^n\,\zeta_{2n+1}\,(2n+1)\Big(\frac{\lambda}{2\pi^2N}\Big)^n\,t^c_{2n,2k}\\[0.5em]
\langle \widetilde{M}^{(2)}\,{W} \rangle_{0;\text{con}} =& -\frac{1}{2}\sum_{n=1}^{\infty}\sum_{k=1}^{\infty}\sum_{\ell=0}^{n}\frac{1}{(2k)!}\left(\frac{\lambda}{2N}\right)^{k}(-1)^n\,\zeta_{2n+1}\,(2n+1)\Big(\frac{\lambda}{8\pi^2\,N}\Big)^n\,\binom{2n}{2\ell}
\,\\[1mm] & \quad\qquad\qquad\times \left(t_{2n-2\ell,2\ell,2k}-t_{2n-2\ell,2\ell}t_{2k}  \right)\, \nonumber,
\end{align}
\label{MWconnected}%
\end{subequations}
with $t^c_{2n,2k}=t_{2n,2k}-t_{2n}t_{2k}$.

Expressing these two correlators in the basis of the $\mathcal{P}_k$ operators \eqref{PK}, after some algebraic manipulations, similar to those described in Appendix \ref{App:Details on N=4 SYM-like theory}, we find
\begin{subequations}
\begin{align}
& \langle \widetilde{M}^{(1)}\,{W} \rangle_{0;\text{con}}=-2\int_0^{\infty}dt\frac{\text{e}^t\,t}{(\text{e}^t-1)^2}\, \mathcal{V}^{(1)}_{2}\, ,
\label{M1Wcon}\\[0.5em]
&  \langle \widetilde{M}^{(2)}\,{W} \rangle_{0;\text{con}} = -2\int_0^{\infty}dt\frac{\text{e}^t\,t}{(\text{e}^t-1)^2}\,
\left( \mathcal{V}^{(2)}_{2}+\mathcal{V}_{3}\right) \, \ ,
\end{align}
\end{subequations}
where
\begin{subequations}
\begin{align}
&\mathcal{V}^{(1)}_{2}= 
\sum_{q=1}^{\infty}\sum_{k=1}^{\infty}(-1)^k\sqrt{q\,k}\,I_{2q}(\sqrt{\lambda})\,J_{2k}\bigg(\frac{t\sqrt{\lambda}}{\pi}\bigg)\,\,\langle \mathcal{P}_{2q}\,\mathcal{P}_{2k} \rangle_0\, \ , 
\label{V12} \\
&\mathcal{V}^{(2)}_2 = 
\sum_{n=1}^{\infty}\sum_{\ell=0}^{n}\sum_{q=1}^{\infty}\sqrt{q}I_{2q}(\sqrt{\lambda})\,(-1)^n\,\Big(\frac{\lambda\,t^2}{8\pi^2N}\Big)^{n}\Big(\frac{N}{2}\Big)^{\ell}\frac{t_{2n-2\ell}}{(2n-2\ell)!}\sum_{i=0}^{\ell-1}\frac{\sqrt{2(2\ell-2i)}}{i!(2\ell-i)!}\, \,\langle \mathcal{P}_{2\ell-2i}\mathcal{P}_{2q} \rangle_0\, \ ,
\label{V22} \\
&\mathcal{V}^{}_3 =  2 \, 
\sum_{q=1}^{\infty}\sum_{p=1}^{\infty}\sum_{k=1}^{\infty}(-1)^{p+q}\sqrt{p\,q\,k}\,J_{2p}\bigg(\frac{t \sqrt{\lambda}}{2 \pi}\bigg)\,J_{2q}\bigg(\frac{t \sqrt{\lambda}}{2 \pi}\bigg)\,I_{2k}(\sqrt{\lambda})\,\,\langle \mathcal{P}_{2p}\mathcal{P}_{2q}\mathcal{P}_{2k} \rangle_0\, . \label{V3}
\end{align}%
\end{subequations}
The large-$N$ expansion of \eqref{V12} can be readily computed by exploiting the large-$N$ expansion \eqref{PPcon} of the correlator $\langle \mathcal{P}_{2q} \mathcal{P}_{2k} \rangle_0$, obtained by solving recursively the Toda equation \eqref{eqToda2ptcon}. 
In the planar limit the sums over $q$ and $k$ can be performed analytically using the identity \eqref{sumq}, while at the sub-leading orders the series over $q$ and $k$ factorize and their sums can be computed 
using the set of mathematical identities collected in Appendix \ref{app:identities}. After a long but straightforward computation, we find 
\begin{align}
\mathcal{V}^{(1)}_{2} =&-\frac{\pi\,t}{2\pi^2+2t^2}\frac{\mathcal{B}(2t)}{2} -\frac{1}{N}\frac{\lambda\,t}{32\pi}\,J_{1}\bigg(\frac{t \sqrt{\lambda}}{\pi}\bigg)\,I_{1}(\sqrt{\lambda}\,) \nonumber \\
& +\frac{1}{N^2}\Bigg[\frac{\lambda\,t}{1536\,\pi^3}\,\left(\lambda\,t^2-\pi^2(\lambda +4)\right) \,I_1\big(\sqrt{\lambda }\,\big)  J_1\bigg(\frac{t \sqrt{\lambda }}{\pi }\bigg) + \frac{\lambda^{3/2}\,t^2}{768 \pi^2}\,I_1\big(\sqrt{\lambda }\big) \,J_0\bigg(\frac{t\sqrt{\lambda}}{\pi }\bigg) \bigg. \nonumber \\
& \Bigg. + \frac{\lambda^2\,t^2}{1536\pi^2}\,I_0\big(\sqrt{\lambda}\,\big) \,J_2\bigg(\frac{t\sqrt{\lambda}}{\pi }\bigg) \Bigg] + O\Big(\frac{1}{N^3}\Big) \, \ ,
\label{LargeNIandJ}
\end{align}
where we introduced the function
\begin{align}
\mathcal{B}(t) \equiv \sqrt{\lambda}\,I_0(\sqrt{\lambda})\,J_1\bigg(\frac{t\sqrt{\lambda}}{2\pi}\bigg) - \frac{t\sqrt{\lambda}}{2\pi}\,I_1(\sqrt{\lambda})\,J_0\bigg(\frac{t\sqrt{\lambda}}{2\pi}\bigg)\, \ .
\label{Bt}
\end{align}

Let us now consider $\mathcal{V}^{(2)}_2$. 
By using the large-$N$ expansion of $t_{2n-2\ell}$ given in \eqref{t2k}, after some algebra we can rewrite \eqref{V22} as
\begin{align}
\mathcal{V}^{(2)}_2 = 4N
\, \mathcal{C}(   \lambda)\,
\sum_{q=1}^{\infty}\sum_{k=1}^{\infty}\sqrt{q\,k}\,(-1)^kI_{2q}(\sqrt{\lambda})J_{2k}\bigg(\frac{t \sqrt{\lambda}}{2 \pi}\bigg)\,\langle \mathcal{P}_{2q}\,\mathcal{P}_{2k} \rangle_0 
\, \ ,
\label{V22a}
\end{align}
where 
\begin{align}
\mathcal{C}(\lambda) &= \frac{4\pi}{t\sqrt{\lambda}}\,J_{1}\bigg(\frac{t \sqrt{\lambda}}{2 \pi}\bigg) + \frac{1}{4N}\Bigg[J_{0}\bigg(\frac{t \sqrt{\lambda}}{2 \pi}\bigg)-1\Bigg] + \frac{1}{N^2}\frac{t^2\lambda}{768\pi^2}\,J_{2}\bigg(\frac{t \sqrt{\lambda}}{2 \pi}\bigg) \nonumber \\
 &-\frac{1}{N^3}\frac{t^3\lambda^{3/2}}{12288\pi^3}\,J_{3}\bigg(\frac{t \sqrt{\lambda}}{2 \pi}\bigg) +\, O\Big(\frac{1}{N^4}\Big)\, \ .
\label{Cexpansion}
\end{align}
The series in \eqref{V22a} is the same as that in \eqref{V12} with $t\to t/2$ and so from \eqref{LargeNIandJ} we get
\begin{align}
\mathcal{V}^{(2)}_2 = &-N\frac{8\pi^2}{4\pi^2+t^2} \,\frac{\mathcal{B}(t)}{\sqrt{\lambda}}\,J_1\bigg(\frac{t\sqrt{\lambda}}{2\pi}  \bigg) + \frac{\pi t}{2(4\pi^2+t^2 )} \, \Bigg[ 1-J_0\bigg(\frac{t\sqrt{\lambda}}{2\pi} \bigg)\Bigg]\,\mathcal{B}(t) \nonumber\\ &-\frac{\sqrt{\lambda}}{4}\,I_1\big(\sqrt{\lambda}\,\big)\,J_1\bigg(\frac{t\sqrt{\lambda}}{2\pi}  \bigg) +\,O\Big(\frac{1}{N} \Big)
\, \ ,
\label{V22b}
\end{align}
Finally, for the last term $\mathcal{V}_3$, we obtain its large-$N$ expansion using the expression 
\eqref{PPPcon} for the correlator $\langle \mathcal{P}_{2p} \mathcal{P}_{2q} \mathcal{P}_{2k} \rangle_0$, together with the identities collected in Appendix \ref{app:identities}, getting
\begin{align}
\mathcal{V}_3 =&\,
\frac{1}{N}\frac{\lambda^{3/2}\,t^2}{64\,\pi^2}\,I_1\big(\sqrt{\lambda }\,\big) J_1\bigg(\frac{t \sqrt{\lambda }}{2\pi}\bigg)^2 -\frac{1}{N^2}\frac{\lambda^2t^2}{512\,\pi^3}\,J_1\bigg(\frac{\sqrt{\lambda } t}{2 \pi }\bigg) \Bigg[t\, I_1\big(\sqrt{\lambda }\,\big) J_0 \bigg(\frac{t \sqrt{\lambda }}{2\pi}\bigg) \Bigg. \nonumber\\
& \Bigg. +\,\pi\, I_2\big(\sqrt{\lambda }\,\big) \,J_1\bigg(\frac{t \sqrt{\lambda}}{2 \pi }\bigg)\Bigg] + \,O\Big(\frac{1}{N^3}\Big)\,  
\, \ .
\label{V3LargeN}
\end{align}
We remark that, although we have explicitly shown only the first terms of large-$N$ expansions, thanks to the Toda equations
it is possible to obtain analytic expressions
for $\mathcal{V}^{(1)}_2$, $\mathcal{V}^{(2)}_2$ and $\mathcal{V}_3$ to any desired order. In fact, we have carried out this computation up to $O(N^{-9})$.

We are now ready to compute the large-$N$ expansion of \eqref{StartN4}. Using the Ansatz
\begin{align}
\widetilde{\mathcal{K}}_N(\lambda)= N\,\widetilde{\mathcal{K}}^{(0)}(\lambda) + \widetilde{\mathcal{K}}^{(1)}(\lambda) + N^{-1}\,\widetilde{\mathcal{K}}^{(2)}(\lambda) + \cdots   \, \ ,
\label{KN4LargeN}
\end{align}
we obtain
\begin{subequations}
\begin{align}
\widetilde{\mathcal{K}}^{(0)}(\lambda) &= \frac{16\pi^2}{\sqrt{\lambda}}\int_0^{\infty}dt\,\frac{\text{e}^{t}\,t}{(1-\text{e}^{t})^2}\,\frac{1}{4\pi^2+t^2}\,J_1\bigg(\frac{t\sqrt{\lambda}}{2\pi}\bigg)\,\mathcal{B}(t)\, \ , \label{K0N4} \\
\widetilde{\mathcal{K}}^{(1)}(\lambda) &=  \frac{\pi}{2}\int_0^{\infty}dt\,\frac{\text{e}^{t}\,t^2}{(1-\text{e}^{t})^2}\frac{1}{\pi^2+t^2}\,\mathcal{B}(2t) + \frac{\sqrt{\lambda}}{2}\, I_1\big(\sqrt{\lambda}\,\big)\int_0^{\infty}dt\,\frac{\text{e}^{t}\,t}{(1-\text{e}^{t})^2}\,J_1\bigg(\frac{t\sqrt{\lambda}}{2\pi}\bigg)^2 \nonumber \\
&-\pi \int_0^{\infty}dt\,\frac{\text{e}^{t}\,t^2}{(1-\text{e}^{t})^2}\frac{1}{4\pi^2+t^2}\,\Bigg[1-J_0\bigg(\frac{t\sqrt{\lambda}}{2\pi}\bigg)\Bigg]\,\mathcal{B}(t)\, \ .
\label{K1N4}
\end{align}
\label{KN40and1}%
\end{subequations}
The expressions of $\widetilde{\mathcal{K}}^{(j)}$ become progressively more complex as 
$j$ increases. They can be found in the ancillary Mathematica file up to $j=8$.

\subsection{The large-\texorpdfstring{$N$}{} expansion of the integrated correlator}
Let us now evaluate the large-$N$ expansion of the integrated correlator $\widetilde{\mathcal{I}}(\lambda)$, which takes the following form
\begin{align}
\widetilde{\mathcal{I}}(\lambda) = \widetilde{\mathcal{I}}^{(0)}(\lambda) + \frac{1}{N}\widetilde{\mathcal{I}}^{(1)}(\lambda) + \frac{1}{N^2}\widetilde{\mathcal{I}}^{(2)}(\lambda) + \cdots ...   \, \ ,
\end{align}
where the coefficients $\widetilde{\mathcal{I}}^{(n)}$ satisfy the recursion relation
\begin{align}
\widetilde{\mathcal{I}}^{(n)}(\lambda) =  \frac{2\,\widetilde{\mathcal{K}}^{(n)}(\lambda)}{ \mathcal{W}^{(0)}_0(\lambda) } -\sum_{i=1}^{n}\frac{ \mathcal{W}^{(i)}_0 (\lambda)}{\mathcal{W}^{(0)}_0(\lambda)}\,\widetilde{\mathcal{I}}^{(n-i)}(\lambda)\, .
\label{I4coefficients}
\end{align}
Using \eqref{w0}, \eqref{W12} and \eqref{KN40and1}, together with the corresponding sub-leading terms, we obtain an exact expression for the 
coefficients $\widetilde{\mathcal{I}}^{(n)}$ valid for any value of the 't Hooft coupling, like for instance\footnote{This result coincides with expression (2.24) of \cite{Pufu:2023vwo}, which corresponds to the planar term of the same observable in the $\mathcal{N}=4$ SYM theory with gauge group $SU(N)$}
\begin{align}
\widetilde{\mathcal{I}}^{(0)}(\lambda)= \frac{8\pi^2}{I_1(\sqrt{\lambda})}\int_0^{\infty}dt\,\frac{\text{e}^{t}\,t}{(1-\text{e}^{t})^2}\,\frac{1}{4\pi^2+t^2}\,J_1\bigg(\frac{t\sqrt{\lambda}}{2\pi}\bigg)\,\mathcal{B}(t)\, \ .    
\label{I0}
\end{align}
The functions $\widetilde{\mathcal{I}}^{(n)}$ can be expanded both at weak and strong coupling.
The weak-coupling expansions are straightforward and, for example, for the first two coefficients they read
\begin{subequations}
\begin{align}
& \widetilde{\mathcal{I}}^{(0)}(\lambda) = \frac{3\lambda^2 \zeta_3}{16\pi^2} + \frac{\lambda^3}{128\pi^4}\,(\pi^2 \zeta_3-25\,\zeta_5) + \frac{\lambda^4}{4096\pi^6}\,(2\pi^4 \zeta_3+30\pi^2\zeta_5+735\,\zeta_7) + \,O(\lambda^5)\, \ , \\[0.5em]
& \widetilde{\mathcal{I}}^{(1)}(\lambda)= \frac{15\lambda^2 \zeta_3}{64\pi^2} -\frac{3\lambda^3}{512\pi^4}\,(2\pi^2 \zeta_3+75\,\zeta_5) + \frac{\lambda^4}{16384\pi^6}\,(14\pi ^4\zeta_3+270\pi^2\zeta_5+9555\,\zeta_7) + O(\lambda^5)\, \ .
\end{align}
\end{subequations}
The strong-coupling expansions, instead, can be obtained with the techniques outlined in Appendix \ref{SC}. After a long computation, we find for the first nine
coefficients the following results
\begin{subequations}
\begin{align}
& \widetilde{\mathcal{I}}^{(0)}(\lambda) \, \underset{\lambda \rightarrow \infty}{\simeq} \,  \lambda^{1/2} + \left(\frac{1}{2}-\frac{\pi^2}{3}\right)   + \frac{3}{8\lambda^{1/2}} + \frac{3}{8 \,\lambda } \,\left(4 \zeta_3+1\right) + O\big(\lambda^{-3/2}\big) \, \ , \\
& \widetilde{\mathcal{I}}^{(1)}(\lambda)  \, \underset{\lambda \rightarrow \infty}{\simeq} \, \frac{\lambda^{1/2}}{8} -\frac{3}{64\,\lambda^{1/2}} -\frac{3}{32\lambda} \left(4 \zeta_3+1\right) -\frac{9}{1024\lambda^{3/2}} \left(64 \zeta_3+21\right)+ O\left(\lambda^{-2}\right)\, \ , \\
&\widetilde{\mathcal{I}}^{(2)}(\lambda)\, \underset{\lambda \rightarrow \infty}{\simeq} \, -\frac{\lambda ^{3/2}}{128} -\frac{9\,\lambda^{1/2}}{1024} -\frac{1}{512}\left(4\zeta_3+1\right) + \frac{9}{16384\,\lambda^{1/2}}\left(176 \zeta_3+9\right) + O\left(\lambda^{-1}\right)\, \ ,  \\
& \widetilde{\mathcal{I}}^{(3)}(\lambda)\, \underset{\lambda \rightarrow \infty}{\simeq} \, \frac{\lambda^{3/2}}{1024} + \frac{11\,\lambda^{1/2}}{8192} + \frac{1}{1024}\left(4\zeta_3+1\right) -\frac{15}{131072\,\lambda^{1/2}}\left(528 \zeta_3-5\right)  +  O\left(\lambda^{-1}\right)\, \ , \\[0.5em]
&  \widetilde{\mathcal{I}}^{(4)}(\lambda)\, \underset{\lambda \rightarrow \infty}{\simeq} \, \frac{\lambda ^3}{184320} + \frac{7 \lambda ^{5/2}}{163840} -\frac{23 \lambda ^2}{61440} + \frac{2483 \lambda ^{3/2}}{3932160} + O(\lambda)\, \ , \\
&  \widetilde{\mathcal{I}}^{(5)}(\lambda)\, \underset{\lambda \rightarrow \infty}{\simeq} \, -\frac{\lambda ^3}{737280} -\frac{21 \lambda ^{5/2}}{1310720} + \frac{23 \lambda ^2}{122880} -\frac{2963 \lambda ^{3/2}}{6291456} +O(\lambda)\, \ , \\
&  \widetilde{\mathcal{I}}^{(6)}(\lambda)\, \underset{\lambda \rightarrow \infty}{\simeq} \, -\frac{\lambda ^{9/2}}{123863040} -\frac{\lambda ^4}{27525120} + \frac{31 \lambda ^{7/2}}{440401920} + \frac{1207 \lambda ^3}{165150720} + O(\lambda^{5/2})\, \ , \\
& \widetilde{\mathcal{I}}^{(7)}(\lambda)\, \underset{\lambda \rightarrow \infty}{\simeq} \, \frac{\lambda ^{9/2}}{330301440}+ \frac{\lambda ^4}{55050240} -\frac{31 \lambda ^{7/2}}{704643072} -\frac{3509 \lambda ^3}{660602880} + O(\lambda^{5/2})\, \ , \\
& \widetilde{\mathcal{I}}^{(8)}(\lambda)\, \underset{\lambda \rightarrow \infty}{\simeq} \, \frac{\lambda ^6}{59454259200} + \frac{\lambda ^{11/2}}{26424115200} -\frac{13 \lambda ^5}{19818086400} -\frac{8747 \lambda ^{9/2}}{3382286745600} + O(\lambda^4)\, \ .
\end{align}
\label{IN4strong}%
\end{subequations}

It is interesting to observe that the leading strong-coupling terms of $\widetilde{\mathcal{I}}^{(2n)}$ are simply related to the corresponding terms written in equation (2.25) of \cite{Pufu:2023vwo}. Indeed, for $n=0,1,2,3$ one can check that
\begin{align}
\widetilde{\mathcal{I}}^{(2n)}(\lambda) = \frac{1}{4^n}\,\mathcal{I}_{\text{W},\text{pert},n}\, \ .    
\end{align}
where in the r.h.s. we used the same notation of \cite{Pufu:2023vwo} denoting by $\mathcal{I}_{\text{W},\text{pert},n}$ the coefficients of the large-$N$ expansion of the integrated correlator for the $\mathcal{N}=4$ SYM theory with gauge group $SU(N)$. However, we did not identify a similar pattern in the sub-leading contributions.

\section{Conclusions}
\label{sec:Conclusions}
In this work, using supersymmetric localization, we studied the integrated correlator \eqref{Wint} in two different theories with gauge group $Sp(N)$ but with a different amount of supersymmetry. Specifically, we derived exact expressions for the coefficients of the large-$N$ expansion of this observable, valid for any value of the 't Hooft coupling, pushing this analysis up to a very high order in $1/N$. The crucial ingredient that allowed us to achieve this is the use of Toda-like equations. 
We then expanded these coefficients at strong coupling. These expansions constitute the main results of this paper and are presented in  
\eqref{IN4strong} for $\mathcal{N}=4$ SYM, and in \eqref{IFINALExpansions} for the $\mathcal{N}=2$ model 

Our findings show that the $\mathcal{N}=4$ SYM theories
with gauge groups $Sp(N)$ and $SU(N)$ are planar equivalent. This is reflected in the fact that the planar term coefficient \eqref{I0} matches the expression (2.24) of \cite{Pufu:2023vwo}. However, the different choice of the gauge group becomes manifest in the different expressions found at the non-planar level. 
Moreover, our results can be regarded as the starting point for a deeper analysis of the properties of the integrated correlators \eqref{Wint}. Specifically, in the context of the maximal supersymmetric theory with gauge group $Sp(N)$, it would be interesting to apply the same computational techniques used in this paper, based on the existence of Toda equations, to investigate the so called ``very strong coupling limit" \cite{Pufu:2023vwo}. Additionally, extending this analysis to incorporate generic dyonic operators and investigating the modular properties of the corresponding integrated correlators, following the approach of \cite{Dorigoni:2024vrb}, would be a valuable direction for further research. Furthermore, for both gauge theories, the expansions we found can be potentially further used to study the resurgence properties of the integrated correlator, along the lines of \cite{Dorigoni:2024dhy,Beccaria:2022kxy}. 

Finally, we observe that the large-$N$ expansions of the correlation functions among the $\mathcal{P}_k$ operators, derived using the Toda equation and presented in Appendix \ref{app:ConnectedCorrelators} as well as in the ancillary Mathematica file, potentially have a wide range of applications extending far beyond the scope of this work. As a matter of fact, any observable can, in principle, be expanded in terms of the $\mathcal{P}_k$ operators, making our results particularly valuable for future investigations within the context of SCFTs with gauge group $Sp(N)$.

\vskip 1cm
\noindent {\large {\bf Acknowledgments}}
\vskip 0.2cm
We are very grateful to Alberto Lerda for carefully reading the manuscript and to Daniele Dorigoni, Marco Billò and Paolo Vallarino for many relevant discussions. 
AP would like to thank the String Theory group at the University of Turin for their hospitality during the initial stages of this work and the High Energy Physics Theory Group at the University of Oviedo for their hospitality during the final stages. Furthermore, MF would like to thank the Theoretical Particle Physics group at SISSA for their hospitality during the final stage of this work. 

This research is partially supported by the MUR PRIN contract 2020KR4KN2 ``String Theory as a bridge between Gauge Theories and Quantum Gravity'' and by
the INFN project ST\&FI
``String Theory \& Fundamental Interactions''.
The work of AP is supported  by the Deutsche Forschungsgemeinschaft (DFG, German Research Foundation) via the Research Grant ``AdS/CFT beyond the classical supergravity paradigm: Strongly coupled gauge theories and black holes” (project number 511311749). 

\vskip 1cm

\appendix

\section{Mathematical identities}
\label{app:identities}
In this appendix we collect the mathematical identities that have been employed for the derivation of the large-$N$ expansion of the integrated correlator $\widetilde{\mathcal{I}}$ in Section \ref{sec:IN4}.

By using the properties of the Bessel function, it can be shown that
\begin{align}
\sum_{q=1}^{\infty}(-1)^q(2q)I_{2q}(\sqrt{\lambda})J_{2q}(x)  = -\frac{\sqrt{\lambda}\,x}{2(x^2+\lambda)}\left(\sqrt{\lambda}\,I_0(\sqrt{\lambda})\,J_1(x)-x\,I_1(\sqrt{\lambda})\,J_0(x)\right)
\label{sumq}
\end{align}
Furthermore, we also consider series of the form
\begin{align}
2\sum_{q=1}^{\infty}q^{2k+1}I_{2q}(\sqrt{\lambda}) 
\label{identityIBessel}
\end{align}
with $k=0,1,2, \cdots$. It is worth noting that, for fixed $k$, the sums over $q$ in the expression above can always be performed. To illustrate this, let's consider the case $k=1$. Upon using the definition of the modified Bessel function, we obtain
\begin{align}
& 2\sum_{q=1}^{\infty}q^3\;I_{2q}(\sqrt{\lambda})= 2 \sum_{q=1}^{\infty}q^3 \sum_{k=0}^{\infty} \frac{1}{k!(k+2q)!}\bigg(\frac{\sqrt{\lambda}}{2}\bigg)^{2k+2q} = 2 \sum_{m=1}^{\infty} \bigg(\frac{\sqrt{\lambda}}{2}\bigg)^{2m} \sum_{q=1}^{m}\frac{q^3}{(m-q)!(m+q)!} \nonumber \\  
&= \sum_{m=1}^{\infty} \bigg(\frac{\sqrt{\lambda}}{2}\bigg)^{2m} \frac{m \;(m+1)}{(m-1)!(m+1)!}= \frac{\lambda}{4}\; I_0(\sqrt{\lambda}) \ ,
\end{align} 
where, in moving from the first to the second line, we have performed the finite sums over $q$ and recognized once again the series expansion of a modified Bessel function. Using this procedure one can resum the series \eqref{identityIBessel} for any value of $k$. For instance, for $k=0,2,3$ it holds that
\begin{subequations}
\begin{align}
& 2\sum_{q=1}^{\infty}q\,I_{2q}(\sqrt{\lambda}) = \frac{\sqrt{\lambda}}{2}\,I_1(\sqrt{\lambda})\, \label{Firstq}\ ,\\
& 2\sum_{q=1}^{\infty}q^5\,I_{2q}(\sqrt{\lambda}) = \frac{\lambda}{4}
\left[\sqrt{\lambda}\,I_1(\sqrt{\lambda})+I_0(\sqrt{\lambda})\right]\, \ ,\\
& 2\sum_{q=1}^{\infty}q^7\,I_{2q}(\sqrt{\lambda}) =\frac{\lambda}{8} \left[\left(3\lambda+2\right) I_0(\sqrt{\lambda})+4\sqrt{\lambda}\, I_1(\sqrt{\lambda})\,\right]\, \ .
\end{align}
\end{subequations}
Exploiting the same strategy one can also resum series of the form
\begin{align}
2\sum_{q=1}^{\infty}(-1)^q\,q^{2k+1}J_{2q}(\sqrt{\lambda})
\label{identityIBesseJ}\,
\end{align}
with $k=0,1,2\cdots$. For example, for the first values of $k$, we obtain
\begin{subequations}
\begin{align}
& 2\sum_{q=1}^{\infty}(-1)^q\,q\,J_{2q}(\sqrt{\lambda}) = -\frac{\sqrt{\lambda}}{2}\,J_1(\sqrt{\lambda})\, \ , \\
& 2\sum_{q=1}^{\infty}(-1)^q\,q^{3}\,J_{2q}(\sqrt{\lambda}) = -\frac{\lambda}{4} \,J_0(\sqrt{\lambda})\, \ , \\
& 2\sum_{q=1}^{\infty}(-1)^q\,q^{5}\,J_{2q}(\sqrt{\lambda}) = \frac{\lambda}{4}\,\left[\sqrt{\lambda}\,
   J_1(\sqrt{\lambda})-J_0(\sqrt{\lambda})\right]\, \ .
\end{align}
\end{subequations}

\section{Details on strong coupling expansions}
\label{SC}
In this Appendix we provide some technical details on the derivation of the strong coupling expansion of integrals of the form 
\begin{align}
   T_{i,n}^{(p)} &= \int_0^{\infty}dt\; \chi_i^{(p)}(t) \; J_n\bigg( \frac{t\sqrt{\lambda}}{2\pi}  \bigg)  \label{Single J}\\
   S_{i,n,m}^{(p)}&=  \int_0^{\infty}dt\; \chi_i^{(p)}(t) \; J_n\bigg( \frac{t\sqrt{\lambda}}{2\pi}  \bigg) \; J_m\bigg( \frac{t\sqrt{\lambda}}{2\pi}  \bigg) \;, \label{Double J}
\end{align}
with $i=1,2,3$ and where
\begin{subequations}
\begin{align}
   & \chi_1^{(p)}(t) = \frac{\text{e}^t \;t^p}{(\text{e}^t+1)^2} \\
   &  \chi_2^{(p)}(t) = \frac{\text{e}^t \;t^p}{(\text{e}^t-1)^2} \\
   &  \chi_3^{(p)}(t) = \frac{\text{e}^t }{(\text{e}^t-1)^2} \frac{t^p}{4\pi^2 + t^2} 
\end{align}
\label{Allchi}
\end{subequations}
are the integration measures we encountered in the main text.

The first step consists in writing the single Bessel function in \eqref{Single J} or the product of two Bessel functions in \eqref{Double J} using their Mellin-Barnes integral representation, that is respectively given by 
\begin{align}
J_n\bigg(\frac{t\sqrt{\lambda}}{2\pi} \bigg) &= \int_{-i \infty}^{i \infty} \frac{ds}{2\pi i} \frac{\Gamma(-s)}{\Gamma(s+ n +1)} \bigg(\frac{t\sqrt{\lambda}}{4\pi}  \bigg)^{2s+ n }  \, ,
      \label{MBJ}
 \\     
J_{n}\bigg(\frac{t\sqrt{\lambda}}{2\pi} \bigg) J_{m}\bigg(\frac{t\sqrt{\lambda}}{2\pi} \bigg) &= \int_{-i\infty}^{i \infty} \frac{ds}{2\pi i} \frac{\Gamma(-s)\Gamma(2s+ 1+ m +n)}{\Gamma(s+ m +1) \Gamma(s+n +1) \Gamma(s+m+n +1)} \bigg( \frac{t\sqrt{\lambda}}{4\pi}   \bigg)^{2s+m+n} .
     \label{MBJJ}
\end{align}
In the case of $T_{i,n}^{(p)}$, substituting  \eqref{MBJ} into \eqref{Single J} we obtain  
\begin{align}
      T_{i,n}^{(p)}=   \int_{-i \infty}^{i \infty} \frac{ds}{2\pi i} \frac{\Gamma(-s)}{\Gamma(s+ n +1)} \bigg(\frac{\sqrt{\lambda}}{4\pi}  \bigg)^{2s+ n } \int_0^{\infty}dt\; \chi_i(t) \;t^{2s+n} \; \;.
     \label{T1}
\end{align}
In the same way, inserting \eqref{MBJJ} in \eqref{Double J}, we find 
\begin{align}
S_{i,n,m}^{(p)}= \int_{-i \infty}^{i \infty} \frac{ds}{2\pi i} \frac{\Gamma(-s)\Gamma(2s+ 1+ m +n)}{\Gamma(s+ m +1) \Gamma(s+n +1) \Gamma(s+m+n +1)} \bigg( \frac{\sqrt{\lambda}}{4\pi}   \bigg)^{2s+m+n}\int_0^{\infty}dt\; \chi_i^{(p)}(t) \;t^{2s+n+m} \;.
\label{T2}
\end{align}
After performing the integral over $t$, we consider the strong coupling limit i.e.,  $\lambda \gg 1$, where both \eqref{T1} and \eqref{T2} receive contributions from poles located on the negative real axis of $s$. In the following, we analyze one by one the possible integration measures \eqref{Allchi} with suitable examples.
\subsection{The integration measure \texorpdfstring{$\chi_1^{(p)}(t)$}{}}
If we consider
\begin{equation}
    \chi_1^{(p)}(t) = \frac{\text{e}^t \;t^p}{(\text{e}^t+1)^2} \;,
\end{equation}
we can perform the integral over $t$ in \eqref{T1} and \eqref{T2} using the integral representation of the Dirichlet $\eta$ function, which reads
\begin{equation}
    \Gamma(z+1)\;\eta(z)= \int_0^{\infty} dt \frac{\text{e}^t \;t^z}{(\text{e}^t+1)^2} \;,
    \label{eta}
\end{equation}
obtaining 
\begin{align}
T_{1,n}^{(p)} &=   \int_{-i \infty}^{i \infty} \frac{ds}{2\pi i} \frac{\Gamma(-s)\Gamma(2s+n+p+1)\eta(2s+n+p)}{\Gamma(s+ n +1)} \bigg(\frac{\sqrt{\lambda}}{4\pi}  \bigg)^{2s+ n }  \, ,\label{T11}\\[0.5em]
S_{1,n,m}^{(p)} &= \int_{-i \infty}^{i \infty} \frac{ds}{2\pi i} \frac{\Gamma(-s)\Gamma(2s+ 1+ m +n)\Gamma(2s+n+m+p+1)\eta(2s+n+m+p)}{\Gamma(s+ m +1) \Gamma(s+n +1) \Gamma(s+m+n +1)} \bigg( \frac{\sqrt{\lambda}}{4\pi}   \bigg)^{2s+m+n}.
\end{align}

As an illustrative example, we consider the strong coupling expansion of $\partial_{\lambda}F_1$ that, from \eqref{F1}, reads
\begin{align}
    \partial_{\lambda}F_1(\lambda)=  -\frac{4}{\lambda}\int_{0}^{\infty}\frac{dt}{t}\frac{\text{e}^t}{(\text{e}^t+1)^2} \; J_2 \bigg(\frac{t \sqrt{\lambda}}{\pi}  \bigg)+ \frac{\log 2}{2\pi^2}\;.
\end{align}
We note that it takes the form of \eqref{Single J} with integration measure $\chi_1^{(-1)}(t)$. Thus, we can use \eqref{T11}, obtaining
\begin{align}
     \partial_{\lambda}F_1(\lambda) = &\frac{\log 2}{2\pi^2} - \frac{4}{\lambda} T_{1,2}^{(-1)} =  \frac{\log 2}{2\pi^2}  -\frac{4}{\lambda} \int \frac{ds}{2\pi i} \frac{\Gamma(-s)\Gamma(2s+2) \eta(2s+1)}{\Gamma(s+3) } \bigg( \frac{\sqrt{\lambda}}{2\pi}  \bigg)^{2s+2} \;. 
\end{align} 
Picking up the residues at $s=-1$ and $s=-2$, which are the only poles located on the negative real axis, we find the exact $\lambda$ dependence of 
    \begin{equation}
    \partial_{\lambda}F_1(\lambda) \underset{\lambda \rightarrow \infty}{\simeq} \,\frac{\log2}{2\pi^2}-\frac{1}{2\lambda} + \frac{\pi^2}{2 \lambda^2} \;.
    \label{SCF1}
\end{equation}

\subsection{The integration measure \texorpdfstring{$\chi_2^{(p)}(t)$}{}}
Now we consider the integration measure 
\begin{equation}
      \chi_2^{(p)}(t) = \frac{\text{e}^t \;t^p}{(\text{e}^t-1)^2} \;.
\end{equation}
As in the previous case, it is possible to perform explicitly the integral over $t$ in \eqref{T1} and \eqref{T2}, this time using the integral representation of the Riemann $\zeta$ function, that reads 
\begin{equation}
    \Gamma(z+1)\;\zeta(z)= \int_0^{\infty} dt \frac{\text{e}^t }{(\text{e}^t-1)^2}t^z \;,
    \label{zeta}
\end{equation}
valid for $\text{Re}(z)>1$, getting
\begin{align}
T_{2,n}^{(p)} &=   \int_{-i \infty}^{i \infty} \frac{ds}{2\pi i} \frac{\Gamma(-s)\Gamma(2s+n+p+1)\zeta(2s+n+p)}{\Gamma(s+ n +1)} \bigg(\frac{\sqrt{\lambda}}{4\pi}  \bigg)^{2s+ n }  \, , \label{T12}\\[0.5em]
S_{2,n,m}^{(p)} &= \int_{-i \infty}^{i \infty} \frac{ds}{2\pi i} \frac{\Gamma(-s)\Gamma(2s+ 1+ m +n)\Gamma(2s+n+m+p+1)\zeta(2s+n+m+p)}{\Gamma(s+ m +1) \Gamma(s+n +1) \Gamma(s+m+n +1)} \bigg( \frac{\sqrt{\lambda}}{4\pi}   \bigg)^{2s+m+n} .
\end{align}
For example, we evaluate the strong coupling expansion of $\partial_{\lambda}\mathcal{M}^{(0)}$ that, from \eqref{M0}, is given by
\begin{equation}
   \partial_{\lambda}\mathcal{M}^{(0)}(\lambda) = \frac{2}{\lambda} \int_{0}^{\infty}dt\;\frac{\text{e}^t\;t}{(\text{e}^t-1)^2} \; J_2 \bigg(\frac{t \sqrt{\lambda}}{\pi}  \bigg) \;= \frac{2}{\lambda}\,T_{2,2}^{(1)}.
\end{equation}
Applying \eqref{T12}, we get 
\begin{align}
    \partial_{\lambda}\mathcal{M}^{(0)}(\lambda) =  \frac{2}{\lambda} \int_{-i \infty}^{i \infty} \frac{ds}{2\pi i} \frac{\Gamma(-s)\Gamma(2s+4)\zeta(2s+3)}{\Gamma(s+ 3)} \bigg(\frac{\sqrt{\lambda}}{4\pi}  \bigg)^{2s+2} \;.
\end{align}
Considering the residues at $s=-1$ and $s=-2$, we finally find the exact form of
\begin{equation}
     \partial_{\lambda}\mathcal{M}^{(0)}(\lambda) \underset{\lambda \rightarrow \infty}{\simeq} \,  \frac{1}{\lambda} - \frac{4\pi^2}{3\lambda^2}\, \ .
     \label{SCM0}
\end{equation}
\subsection{The integration measure \texorpdfstring{$\chi_3^{(p)}(t)$}{}}
The last integration measure to be considered is
\begin{equation}
    \chi_3^{(p)}(t) = \frac{\text{e}^t \;}{(\text{e}^t-1)^2} \frac{t^p}{4\pi^2 + t^2} \;.
\end{equation}
This time, the integral over $t$ in \eqref{T1} and \eqref{T2} can not be performed explicitly as in the previous two cases. However,  following \cite{Pufu:2023vwo}, we introduce the function 
\begin{equation}
    \mathcal{J}(s)= \int_0^{\infty}dt  \frac{\text{e}^t \;}{(\text{e}^t-1)^2} \,\frac{t^{2s+3}}{4\pi^2 + t^2}\;,
    \label{J}
\end{equation} which is
defined for $\text{Re}(s)>-1$, but can be analytically extended to the entire complex plane. This turned out to be possible by repeatedly using the following recurrence relation
\begin{equation}
      \mathcal{J}(s) + 4\pi^2 \mathcal{J}(s-1)= \int_0^{\infty} dt\frac{\text{e}^t\; }{\left(\text{e}^t-1\right)^2} t^{2 s+1}=  \Gamma(2s+2)\zeta(2s+1) \;.
      \label{Ricorrenza}
\end{equation}
The analytically extended function has simple poles at negative integer values. Then, by using \eqref{J}, the expressions \eqref{T1} and \eqref{T2} can be rewritten as
\begin{align}
T_{3,n}^{(p)} &=   \int_{-i \infty}^{i \infty} \frac{ds}{2\pi i} \frac{\Gamma(-s)}{\Gamma(s+ n +1)} \, \mathcal{J}\bigg( s+\frac{n+p-3}{2} \bigg) \,\bigg(\frac{\sqrt{\lambda}}{4\pi}  \bigg)^{2s+ n } 
\; , \label{T13}\\[0.5em]
S_{3,n,m}^{(p)} &= \int_{-i \infty}^{i \infty} \frac{ds}{2\pi i} \frac{\Gamma(-s)\Gamma(2s+ 1+ m +n)}{\Gamma(s+ m +1) \Gamma(s+n +1) \Gamma(s+m+n +1)} \,
\mathcal{J}\bigg(s + \frac{n+m+p-3}{2} \bigg)\, \bigg( \frac{\sqrt{\lambda}}{4\pi}\bigg)^{2s+m+n}\,. 
    \label{T32} 
\end{align}

In the strong coupling limit, the function $\mathcal{J}(s)$ contributes in two distinct ways. First, with its values at negative half-integer values, that could turn out to be poles for the remaining part of the integrand. Second, with its residues at negative integers values.

In particular, the values that $\mathcal{J}(s)$ assumes at negative half-integer numbers can be deduced iteratively from \eqref{Ricorrenza} by starting with 
\begin{equation}
    \mathcal{J}\bigg( \frac{1}{2}\bigg) =  \frac{ \pi ^2}{3} \,(10-\pi ^2) \;.
\end{equation}
In this way, we get, for example
\begin{equation}
   \mathcal{J}\bigg(\!\!-\frac{1}{2}\bigg) = \frac{\pi ^2-9}{12} \,
   \;, \quad   \mathcal{J}\bigg(\!\!- \frac{3}{2}\bigg)= \frac{3-\pi ^2}{48 \,\pi ^2} \;, \quad \mathcal{J}\bigg(\!\!- \frac{5}{2}\bigg)= \frac{12 \zeta_3 +\pi ^2-3}{192\, \pi ^4} \;.
   \label{Value J}
\end{equation}
Moreover, from \eqref{Ricorrenza}, we find out a recurrence relation for the residues of $\mathcal{J}(s)$ in $s=-m$, 
$m \in \,\mathbb N$,
that reads
\begin{equation}
    \text{Res}\; \mathcal{J}(s) \big|_{s=-m} + 4\pi^2 \; \text{Res} \;\mathcal{J}(s) \big|_{s=-m-1}  = (-1)^{m+1} \frac{(1-2m) \, \zeta_{2m}}{(2\pi)^{2m}} \;.
    \label{Recurrence 2}
\end{equation}
Since $\mathcal{J}(s)$ is regular in $s=0$, we deduce that 
\begin{equation}
     \text{Res}\; \mathcal{J}(s) \big|_{s=0}=0 \;,
\end{equation}
and from this initial value, exploiting \eqref{Recurrence 2}, we can find the residues in all the negative integers.

As an example, we explicitly evaluate the strong coupling expansion of $\widetilde{\mathcal{I}}^{(0)}$, defined in \eqref{I0}, which is given by 
\begin{equation}
\widetilde{\mathcal{I}}^{(0)}(\lambda) =  \frac{8\pi^2}{I_1(\sqrt{\lambda})} \int_0^{\infty}dt\,\frac{\text{e}^{t}\,t}{(1-\text{e}^{t})^2}\,\frac{1}{4\pi^2+t^2}\,J_1\bigg(\frac{t\sqrt{\lambda}}{2\pi}\bigg)\,\mathcal{B}(t)\, \ .
\label{I0startN4}
\end{equation}
We start by observing that since the function $\mathcal{B}(t)$, defined in \eqref{Bt}, is the sum of two terms, the expression \eqref{I0startN4} can then be rewritten as 
\begin{equation}
\widetilde{\mathcal{I}}^{(0)}(\lambda) = \widetilde{\mathcal{I}}^{(0)}_{a}(\lambda)+\widetilde{\mathcal{I}}^{(0)}_{b}(\lambda)
\end{equation}
with
\begin{align}
    & \widetilde{\mathcal{I}}^{(0)}_{a}(\lambda)= 8\pi^2\,\frac{\sqrt{\lambda}\,I_0(\sqrt{\lambda})}{I_1(\sqrt{\lambda})}\int_0^{\infty}dt\;\frac{\text{e}^t\;t}{(\text{e}^t-1)^2} \frac{1}{4\pi^2+t^2} \; J_1\bigg( \frac{t\sqrt{\lambda}}{2\pi}\bigg)\,J_1 \bigg(  \frac{t\sqrt{\lambda}}{2\pi}  \bigg)\, \ , \\[0.5em]
    & \widetilde{\mathcal{I}}^{(0)}_{b}(\lambda)= - 4\pi\sqrt{\lambda}\int_0^{\infty}dt\;\frac{\text{e}^t\;t^2}{(\text{e}^t-1)^2} \frac{1}{4\pi^2+t^2}  \;J_1\bigg( \frac{t\sqrt{\lambda}}{2\pi}\bigg)\,J_0 \bigg(  \frac{t\sqrt{\lambda}}{2\pi}  \bigg) \;.
\end{align}
We recognize that both the previous expressions take the form of \eqref{T2}, with integration measures given by $\chi_3^{(1)}(t)$ and $\chi_3^{(2)}(t)$ respectively. Then, using \eqref{T32}, we can rewrite them as 
\begin{align}
    & \widetilde{\mathcal{I}}^{(0)}_{a}(\lambda)= 8\pi^2\,\frac{\sqrt{\lambda}\,I_0(\sqrt{\lambda})}{I_1(\sqrt{\lambda})} \int_{-i \infty}^{i \infty} \frac{ds}{2\pi i} \,\frac{\Gamma(-s)\Gamma(2s+ 3)}{\Gamma(s+2)^2 \Gamma(s+3)} \mathcal{J}(s)\,
    \bigg(\frac{\sqrt{\lambda}}{4\pi}\bigg)^{2s+2} 
    \, \ , \label{Ka}\\[0.5em]
    &  \widetilde{\mathcal{I}}^{(0)}_{b}(\lambda)= - 4\pi\sqrt{\lambda} \int_{-i\infty}^{i \infty}\frac{ds}{2\pi i} \,\frac{\Gamma(-s)\Gamma(2s+2)}{\Gamma(s+2)^2\Gamma(s+1)}\,\mathcal{J}(s)\,  \bigg( \frac{\sqrt{\lambda}}{4\pi}\bigg)^{2s+1}  
    \;. \label{Kb}
\end{align}
At strong coupling, both \eqref{Ka} and \eqref{Kb} receive contributions from the pole of $\mathcal{J}(s)$ located at $s=-1$, as well as from the poles of the ratio of the Bessel functions at the negative half integers,  which in turn require us to employ the expressions \eqref{Value J}. Considering both these contributions we finally obtain
\begin{equation}
\widetilde{\mathcal{I}}^{(0)}  \, \underset{\lambda \rightarrow \infty}{\simeq} \,  \sqrt{\lambda} + \left(\frac{1}{2}-\frac{\pi^2}{3}\right)   + \frac{3}{8\sqrt{\lambda}} + \frac{3}{8 \,\lambda } \,(1+ 12 \zeta_3) + O\left(\lambda^{-3/2}\right) \;.
\end{equation}

\section{Details on the Gaussian theory}
\label{App:Details on N=4 SYM-like theory}
In this Appendix, we provide further details on the computations related to the application of the Toda equation in the Gaussian matrix model.
\subsection{Large-\texorpdfstring{$N$}{} expansion of the connected correlator}
\label{App:Large N expansion of the connected correlator}
In this Appendix, we derive the expression for $\mathcal {M}_0^{(0)}$ given in \eqref{M0}. 
We recall that the operator ${M}$ is defined in \eqref{Msp} as follows
\begin{equation}
     {M} = -\sum_{n=1}^{\infty}(-1)^n(2n+1)\zeta_{2n+1}\left(\frac{\lambda}{8\pi^2N}\right)^n\,\text{tr}a^{2n}\, \ .
     \nonumber
\end{equation}
To obtain the leading coefficient of the large-$N$ expansion of its v.e.v in the Gaussian matrix model, we insert \eqref{t2k} into the v.e.v. of \eqref{Msp}, obtaining
\begin{align}
\mathcal{M}^{(0)}_0(\lambda)= & -2 \sum_{n=1}^{\infty} (-1)^n  \frac{(2n+1)! \,\zeta_{2n+1}}{n!(n+1)!} \bigg( \frac{\lambda}{16\pi^2}  \bigg)^n  
    \nonumber \\
    & =  - 2 \int_0^{\infty} dt \frac{\text{e}^t \; t}{(\text{e}^t-1)^2} \sum_{n=1}^{\infty} \frac{(-1)^n}{n!(n+1)!} \bigg( \frac{t\sqrt{\lambda}}{4\pi}  \bigg)^{2n} 
    \nonumber\\
    &=  - 2 \int_0^{\infty} dt \frac{\text{e}^t\;t}{(\text{e}^t-1)^2}\bigg[\frac{4\pi}{\sqrt{\lambda}t}J_1 \bigg( \frac{t\sqrt{\lambda}}{2\pi}   \bigg) -1  \bigg] \; .\label{MO4}
\end{align}

\subsection{Toda equation for the 3-point function \texorpdfstring{$t^{(N)\,c}_{2k_1,2k_2,2k_3}$}{} and \texorpdfstring{$\langle \mathcal{P}_{2n_1}\mathcal{P}_{2n_2}\mathcal{P}_{2n_3} \rangle_0$}{}}
In this Appendix we explain how to determine the large-$N$ expansion of the three-point correlator $\langle \mathcal{P}_{2n_1}\mathcal{P}_{2n_2}\mathcal{P}_{2n_3} \rangle_0$. As a first step, using the definition \eqref{PK}, we write it as
\begin{align}
\langle \mathcal{P}_{2n_1}\mathcal{P}_{2n_2}\mathcal{P}_{2n_3} \rangle_0 = \left[\prod_{i=1}^{3}\sqrt{n_i}\sum_{\ell_i=0}^{n_i-1}(-1)^{\ell_i}\left(\frac{N}{2}\right)^{\ell_i-n_i}\frac{(2n_i-\ell_i-1)!}{(\ell_i)!\,(2n_i-2\ell_i)!}\right]t^c_{2n_1-2\ell_1,2n_2-2\ell_2,2n_3-2\ell_3}\, \ .
\label{3Pconnected}
\end{align}
Then, using the method explained in Section \ref{subsec:Todafort} we derive the Toda equation for the connected function $t^{(N)\,c}_{2k_1,2k_2,2k_3}$, namely
\begin{align}
& (k_1+k_2+k_3+1)(k_1+k_2+k_3)t^{(N)\,c}_{2k_1,2k_2,2k_3} =
\frac{N(2N+1)}{2}\,\Big[t^{(N+1)\,c}_{2k_1,2k_2,2k_3}+t^{(N-1)\,c}_{2k_1,2k_2,2k_3}-2\,t^{(N)\,c}_{2k_1,2k_2,2k_3}\Big] \nonumber\\
& \quad\quad\quad\quad\quad+ 
\frac{2}{N(2N+1)}
\sum_{\sigma \in \mathcal{Q}}k_{\sigma(3)}(k_{\sigma(3)}+1)(k_{\sigma(1)}+k_{\sigma(2)}+1)(k_{\sigma(1)}+k_{\sigma(2)})\,t^{(N)\,c}_{2k_{\sigma(1)},2k_{\sigma(2)}}\,t^{(N)}_{2k_{\sigma(3)}} \nonumber\\
& \quad\quad\quad\quad\quad 
-\frac{8}{N^2(2N+1)^2}
\prod_{i=1}^{3}k_i(k_i+1)t^{(N)}_{2k_i}\, \ ,
\label{eqToda3ptcon}
\end{align}
where we introduce the permutation set 
\begin{align}
\mathcal{Q} \equiv \{(1,2,3),\,(3,1,2),\,(2,3,1)\}\, \ ,    
\end{align}
while the functions $t^{(N)\,c}_{2k}$ and $t^{(N)\,c}_{2k_i,2k_j}$ are determined through the equation \eqref{TodaTracesN41pt} and the equation \eqref{eqToda2ptcon}, respectively. 

The equation \eqref{eqToda3ptcon} holds for any value of 
$N$. In the following, we focus on determining its solution in the large-$N$ limit. To this end, we use the following ansatz
 \begin{align}
t_{2k_1,2k_2,2k_3}^{(N)\,c} =  \sum_{q=1}^{k_1+k_2+k_3-1}d_{\text{3pt};q}(k_1,k_2,k_3)N^{k_1+k_2+k_3-q}\, \ .
\label{t3LargeN}
 \end{align}
We find that the coefficients $d_{\text{3pt};q}(k_1,k_2,k_3)$ take the following form
\begin{align}
d_{\text{3pt};q}(k_1,k_2,k_3) = \left[\prod_{i=1}^{3}2k_i(2k_i+2)d_{\text{1pt};1}(k_i)\right]\widetilde{d}_{\text{3pt};q}(2k_1,2k_2,2k_3)\, \ , 
\end{align}
where the coefficient $d_{\text{1pt};1}(k)$ is given in \eqref{d1pt} while the coefficients $\widetilde{d}_{\text{3pt};q}(2k_1,2k_2,2k_3)$ can be determined iteratively for any $q$. For example, for $q=1,2$, they read
\begin{align}
\widetilde{d}_{\text{3pt};1}(k_1,k_2,k_3) = \frac{1}{256}, \qquad   \widetilde{d}_{\text{3pt};2}(k_1,k_2,k_3) =  \frac{k_1+k_2+k_3-2}{2048}\, \ .
\end{align}
As a final step we insert \eqref{t3LargeN} into \eqref{3Pconnected}. This way we obtain
\begin{align}
\langle \mathcal{P}_{2n_1}\,\mathcal{P}_{2n_2}\,\mathcal{P}_{2n_3} \rangle_0 &=  512\,\left[\prod_{i=1}^{3}\sqrt{n_i}\sum_{\ell_i=0}^{n_i-1}\frac{(-1)^{\ell_i}(2n_i-\ell_i-1)!}{(\ell_i)!(n_i-\ell_i)!(n_i-\ell_i-1)!}\right]\times \nonumber\\
& \sum_{q=1}^{n_1+n_2+n_3-\ell_1-\ell_2-\ell_3-1}\widetilde{d}_{\text{3pt};q}(2n_1-2\ell_1,2n_2-2\ell_2,2n_3-2\ell_3)N^{-q}\, \ .    
\end{align}
Given the explicit expressions for the coefficients $\widetilde{d}_{\text{3pt};q}(k_1,k_2,k_3)$ and using the set of mathematical identities \eqref{gMath}, we can systematically derive the large-$N$ expansion of \eqref{3Pconnected} to any desired order. For the purposes of this work, it was sufficient to compute up to $O(N^{-8})$. The first terms of the expansion are provided in Appendix \ref{app:ConnectedCorrelators}.

\subsection{Large-\texorpdfstring{$N$}{} expansions of 2- and 3-point correlators in \texorpdfstring{$\mathcal{N}=4$}{} SYM}
\label{app:ConnectedCorrelators}
Here we collect the large $N$ expansions up to the order $O(N^{-3})$  of the 2- and 3-point correlators among the $\mathcal{P}_{k}$ operators \eqref{PK} in $\mathcal{N}=4$ SYM. 
\begin{align}
 \big\langle \mathcal{P}_{2n_1}\mathcal{P}_{2n_2} \big\rangle_{0} &= \delta_{n_1 n_2} + \frac{\sqrt{n_1 n_2}}{2\,N} + \frac{\sqrt{n_1\, n_2}}{48\,N^2}  (n_1^2+n_2^2-2)\left(n_1^2+n_2^2-1\right) 
 \nonumber \\
& + \frac{\sqrt{n_1 n_2}}{576\,N^3}\,\Big[n_1^6+6\, n_2^2
   n_1^4-11 \,n_1^4+6\, n_2^4 n_1^2-36\, n_2^2 n_1^2+34\,
   n_1^2+n_2^6-11\, n_2^4+34\, n_2^2-24\Big]
   \nonumber \\
   &+ O \Big(\frac{1}{N^4}
   \Big)\, \ ,
\label{PPcon}   
\\
\nonumber \\
 \big\langle \mathcal{P}_{2n_1}\mathcal{P}_{2n_2}\mathcal{P}_{2n_3}  \big\rangle_{0} &= \frac{2 \,\sqrt{n_1 n_2 n_3}}{N} + \frac{\sqrt{n_1 n_2 n_3}}{2\,N^2} \,
   \left(n_1^2+n_2^2+n_3^2-1\right) +\frac{\sqrt{n_1 n_2 n_3}}{144\,N^3} \,\Big[ n_1^6 +6\, (n_2^2+n_3^2) n_1^4
   \Big. 
   \nonumber \\
&   \Big. 
   +12\, n_2^2 n_3^2
   n_1^2+n_2^6+n_3^6+6 \left(n_1^2+n_2^2\right) n_3^4+6
   \left(n_1^2+n_3^2\right)n_2^4 +19
   \left(n_1^2+n_2^2+n_3^2\right) \Big. \nonumber \\
& \Big.  -24 \left(n_1^2
   n_2^2+n_3^2 n_2^2+n_1^2 n_3^2\right)-8\left(n_1^4+n_2^4+n_3^4\right)-12\Big]+ 
   O \Big(\frac{1}{N^4}\Big)\, \ .
\label{PPPcon}   
\end{align}
The full set of coefficients up to the order  up to $O(N^{-8})$ has been compiled in the ancillary Mathematica file.

\printbibliography

\end{document}